\documentclass[prl]{revtex4}
\usepackage{graphicx}
\usepackage{amsmath,amsthm,amssymb}
\newcommand{\be}{\begin{equation}}
\newcommand{\ee}{\end{equation}}

\newcommand{\EE}{\check E}
\newcommand{\fl}{d}
\newcommand{\vlr}{v_{LR}}
\newcommand{\In}{\cal I}
\newcommand{\Out}{\cal O}
\newcommand{\gub}{\tau}
\newcommand{\sgn}{{\rm sign}}
\newcommand{\Slow}{{\cal SLOW}}
\newcommand{\Fast}{{\cal FAST}}
\newcommand{\B}{{\cal B}}
\newcommand{\con}{c_{loc}}
\newtheorem{definition}{Definition}
\newtheorem{corollary}{Corollary}
\newtheorem{lemma}{Lemma}
\newtheorem{theorem}{Theorem}

\begin{document}

\title{Quasi-adiabatic Continuation for Disordered Systems: Applications to Correlations, Lieb-Schultz-Mattis, and Hall Conductance}
\author{M. B. Hastings}
\email{mahastin@microsoft.com}
\affiliation{Microsoft Research Station Q, CNSI Building, University of California, Santa Barbara, CA,
93106, USA}

\begin{abstract}
We present a possible definition of a mobility gap for a many-body quantum system, in analogy to definitions of dynamical localization for
single particle systems.  Using this definition, we construct ``corrected" quasi-adiabatic continuation operators.  We show that these
operators have the same locality properties as the ordinary quasi-adiabatic continuation operators
do in the case of a spectral gap, and that they approximate adiabatic evolution in the
region with a mobility gap just as the ordinary  operators do with a spectral gap.  Further, under an appropriate definition of a unique
ground state (equivalently, an absence of topological order as defined in the
text),
we show how to introduce virtual fluxes and prove bounds similar to those obtained on an energy for the effect of
inserting $2\pi$-flux.  Armed with these results, we can directly carry over previous results proven in the case of a spectral gap.
We present a statement of a higher-dimensional Lieb-Schultz-Mattis theorem for disordered systems (however, the lack of translational
invariance presents us from proving the vanishing of the gap but rather only lets us prove a weaker statement that either the gap
becomes superpolynomially small or the expectation value of the flux insertion operator varies in a particular way); we present a proof of decay of
correlation functions; and we present a proof of Hall conductance quantization
under very mild density-of-states assumptions defined later.  We also generalize these definitions to the
case of a ``bulk mobility gap", in the case of a system with boundaries, and present a proof of Hall conductance quantization on an annulus
under appropriate assumptions.

Further, we present a new ``optimized" quasi-adiabatic continuation operator which simplifies previous estimates and tightens bounds in certain cases.  This is presented in an appendix which can be read independently of the rest of the paper as it also improves estimates
in the case of systems with a spectral gap.  
This filter function used decays in time at least as fast as
${\cal O}(\exp(-t^\alpha))$ for all $\alpha<1$, a class of decay that is called ``subexponential" (a more precise and tighter description of
what is possible  is below).  Using this function it is possible to tighten recent estimates of the Hall conductance quantization for
gapped systems\cite{hall} to an error which also decays subexponentially (again, more precise descriptions are below), rather than
just as an exponential of a power.
\end{abstract}
\maketitle

\section{Introduction}
Recently, the use of Lieb-Robinson\cite{lr1,lr2,lr3} bounds, combined with appropriately chosen filter functions has led to significant
progress in proving results about quantum many-body systems.  Examples include the higher-dimensional Lieb-Schultz-Mattis theorem\cite{lsm},
where this combination of techniques was introduced, decay of correlation functions in gapped systems\cite{lsm,net,lr2,lr3}, an area law for entanglement
entropy for arbitrary one dimensional gapped systems\cite{1darea}, a
simpler proof of Goldstone's theorem for gapped Hamiltonians\cite{berry1}, and, most recently, a proof of Hall conductance quantization for
interacting electrons without averaging assumptions\cite{hall}.

However, these results suffer from one major limitation: they require a spectral gap.  However, in many cases we would prefer to require,
instead, the weaker requirement of a {\it mobility gap}: a gap to propagating excitations.  For non-interacting systems, the concept of localization has been
around since Anderson's early work\cite{local}.  Recently, interest has arisen in the possibility of localization in interesting
systems.  Interesting results include the possibility of a many-body localization transition\cite{loc}, and theorems proving
many-body localization for certain interacting systems, albeit in
a special case that can be mapped to a non-interacting system\cite{tjo}.  In the single
particle case, one can define localization in different ways.  One way is in terms of the properties of the single particle
eigenstates, while a different way is in terms of the dynamics\cite{dloc}.  For a many-body system, the concept of single particle
eigenstates no longer makes sense.  Thus, we need to seek another definition.

In this paper, we present a possible definition of the concept of a mobility gap.  Further definitions are required to specify how the
ground state should be distinguished from other states (definition \ref{unique} below), and to generalize these concepts to
open boundary conditions with gapless edge modes, as would appear in a Hall system.
Using the appropriate definitions, we show how to generalize the concept of a quasi-adiabatic continuation operator\cite{lsm,hwen} to this kind of
system, while preserving the needed locality properties.  Given these results, we then are able to directly carry over many
of the results previously shown using spectral gap.  Under appropriate assumptions, we prove exponential decay of correlation functions,
a version of a higher-dimensional Lieb-Schultz-Mattis theorem, and Hall conductance quantization.

Our mobility gap definition describes an assumption that the propagation of low energy excitations is very slow.  This assumption
is stronger than the usual Lieb-Robinson bound.  Lieb-Robinson bounds hold for very general classes of Hamiltonians (essentially,
any lattice Hamiltonian with short-range interactions and a bound on the interaction strength).  These bounds were introduced in \cite{lr1}.
In \cite{lsm}, the idea of shifting certain terms in the equation of motion in a way that maintained the norm
was introduced to show that these bounds hold in a way that is independent of the dimension
of the Hilbert  space on each site.  In \cite{lr3}, a more general description on arbitrary lattices was given, albeit with dimension-dependent bounds, and in \cite{lr2}, the 
dimension-independent bounds were presented for arbitrary lattices and an extension to interactions decaying slower than
exponential was given (this extension will be used in the appendix of this paper).

In this paper,
we also present a more general definition of a quasi-adiabatic continuation operator, that contains the previous definitions.  Further,
we define an exact quasi-adiabatic continuation operator in the appendix with improved decay properties in time (exactly matching adiabatic
evolution while also decaying in time as an exponential of a polynomial of the time).  The use of this operator significantly
simplifies the error estimates, in particular in cases where we need a Lieb-Robinson bound for quasi-adiabatic evolution.  For example,
it significantly tightens the error estimates in the recent proof of Hall conductance quantization.
The appendix can be read separately.

\section{Definitions of a Mobility Gap}
We consider lattice Hamiltonians of the following form:
we assume that $H$ is a sum of terms
\be
H_0=\sum_Z H_Z,
\ee
where each $H_Z$ is supported on set $Z$, and obeys the following.
First, the diameter of every set $Z$ is at most $R$.  Second,
\be
{\rm sup}_{i} \sum_{Z \ni i} \Vert H_Z \Vert \leq J,
\ee
where the supremum is over sites $i$.  We let $V$ denote the number of sites in the system.  Thus, $\Vert H_0 \Vert \leq JV$.
We refer to $R$ as the ``range" and $J$ as the ``interaction strength".

We use ${\rm dist}(\cdot,\cdot)$ for a metric on the lattice; we measure distances between pairs of sites, pairs of sets, or a site
and a set using the same function.  The distance between a pair of sets is defined to be the minimum over pairs of sites in the pair of
sets of the distance between the sites, and similarly for the distance between a set and a site.  We use ${\rm diam}(\cdot)$ to
indicate the diameter of a set.  For any set $A$, we use $\overline A$ to denote the complement of $A$.  We use $L$ to denote various
measures of the linear size of the system: for the case of a torus later, for example, we will consider an $L$-by-$L$ torus so that
$V=L^2$.

We use $\Psi_0$ to indicate the ground state of $H_0$, and similarly we use $P_0=|\Psi_0\rangle\langle \Psi_0|$ to indicate the
projector onto this ground state.  We use $E_0$ to denote the energy of state $\Psi_0$.
We use $\Vert \cdot \Vert$ to denote the operator norm and $\Vert \cdot \Vert_1$ to denote
the trace norm and we use $| \cdot |$ to denote the $l^2$ norm of a vector.

We use $C$ to refer to numeric constants of order unity.  If we need multiple constants in the same expression, we use
$C_1,C_2,...$.  We use ${\rm poly}(...)$ to refer to quantities bounded by a polynomial
in their arguments.
We use ``computer science" big-O notation: that is, indicating that a quantity is ${\cal O}(x)$ indicates that
it is bounded by a constant times $x$ for sufficiently large $x$.
We use $\exp(-{\rm poly}(L))$ to indicate that a quantity is ${\cal O}(\exp(-L^{\alpha})$, for
some $\alpha>0$.  When we express bounds in term of the quantities $L,\lambda_{min},t_{max},\gub$, this is always at fixed
value of the quantities $J/\gamma,R,\con,\xi$ (these quantities are defined below), and we use
$c$ to denote quantities which may depend on $J/\gamma,R,\con,\xi$.  
That is, if we state that a quantity is bounded by $(J/\lambda_{min}) \exp(-cL)$, we mean that the constant $c$ is positive but may depend
on $J/\gamma,R,\con,\xi$.  When we state that a quantity is ``superpolynomially small", we mean that it is superpolynomially
small in $L$, for fixed
$J/\gamma,R,\con,\xi$; in all such cases where we use the term ``superpolynomially small", we assume (and explicitly state) a
polynomial dependence of quantities 
$\lambda_{min},\gub$ on $L$ and a superpolynomially dependence of $t_{max}$ on $L$.

Before the definitions, some discussion is in order regarding ``filter functions".  These functions play an essential role in the
application of Lieb-Robinson bounds to many-body systems.  The combination of these functions with Lieb-Robinson bounds was
introduced in \cite{lsm}.  Broadly speaking, there are many places where, for a given operator $O$, we would like to construct
a state $\tilde f(H_0-E_0) O |\Psi_0\rangle$, where $\tilde f(H_0-E_0)$ is some function of the Hamiltonian $H_0$.  That is, if $H_0$ has eigenvectors $\Psi_i$ with
corresponding eigenvalues $E_i$, then $\tilde f(H_0-E_0)$ has the same eigenvectors but has the eigenvalues $\tilde f(E_i-E_0)$.  In many such cases,
the function $\tilde f(\omega)$ that we would like to construct is not smooth near $\omega=0$.  The two functions that we would most
like to construct are the step function and the function $1/\omega$, which are used in proving correlation decay and in
defining quasi-adiabatic continuation, respectively.  If a system has an energy gap, then we can define a smooth filter function,
$\tilde f'(\omega)$, with the property that $\tilde f'(\omega)$ is smooth and such that $\tilde f(\omega)-\tilde f'(\omega)$ is small for $|\omega|$ larger
than the energy gap.  The smoothness property is used to show that the Fourier transform of $\tilde f'$ is rapidly decaying in time,
and hence to approximate $\tilde f(H_0-E_0) O |\Psi_0\rangle=\int{\rm d}t f(t) \exp(i H_0 t) O \exp(-i H_0 t)| \Psi_0 \rangle$, by
a local operator acting on $\Psi_0$, using Lieb-Robinson bounds to show locality of
$\exp(i H_0 t) O \exp(-i H_0 t)$ for fixed time.  The smallness of 
$\tilde f(\omega)-\tilde f'(\omega)$ for $\omega$ larger than the energy gap suffices to show that
$\left|\tilde f(H_0-E_0) O |\Psi_0\rangle-f(H_0-E_0) O |\Psi_0 \rangle\right|$ is small.
There have been two main classes of filter functions considered.  One can consider filter functions which decay exponentially in
time at the cost of an exponentially small error in $\tilde f(\omega)-\tilde f'(\omega)$.  These functions, which
we will call ``Gaussian filters" (they are not equal to Gaussians, but have similar decay properties) often give the best bounds.  The other class is filter functions was first considered by Osborne in \cite{tjoq} as a modification of the Gaussian idea.
We will call these functions
``exact filters".  Exact filters have 
$\tilde f(\omega)-\tilde f'(\omega)$ identically equal to zero
for $|\omega|$ larger than the gap.  These functions are easier to work with, but they often
give bounds that decay only faster than any power (in the appendix, we present a construction of these filter functions that
leads to the Fourier transform of $\tilde f$ decaying as an exponential of a polynomial in time, for polynomial arbitrarily close to linear, but  in the main text we content
ourselves with superpolynomial decay); they also make it especially easy to
prove Lieb-Robinson bounds for evolution under quasi-adiabatic continuation.
In this paper, we will consider many of the definitions in generality, using
abstract fiter functions.  This will enable us to either find tighter bounds, or to simplify the proofs, depending on preference.

Using such filter functions, one can define a quasi-adiabatic continuation operator to be an
operator
\be
\label{uncorr}
i{\cal D}(H_s,\partial_s H_s)=\int {\rm d}t F(\gamma t) \exp(i H_s t) (\partial_s H_s) \exp(-i H_s t),
\ee
where $F(t)$ is some filter function such that its Fourier transform
$\tilde F(\omega)$ approximates $-1/\omega$ for $|\omega|\geq 1$,
$F(t)$ decays rapidly in time, and $F(t)$ is odd in time and $\tilde F(0)=0$.
Then, the Fourier transform of $F(\gamma t)$ approximates $-1/\omega$ for $|\omega| \geq \gamma$ and so we
approximate adiabatic evolution given a spectral gap $\gamma$.
Later, we modify this definition to account for a mobility gap.  

First, some definitions:
\begin{definition}
For any set $A$, we define $b_l(A)$ to be the set of sites within distance $l$ of set $A$.
\end{definition}

We use a Lieb-Robinson bound in the following form:
\begin{lemma}
Given any operator $O$ supported on a set $A$, for any $l$ and any $t$ with $|t|\leq l/\vlr$, the operator
\be
O(t)\equiv \exp(i H_0 t) O \exp(-i H_0 t)
\ee
can be approximated by an operator $O_l(t)$ supported on $b_l(A)$ up to an error
\be
\Vert O_l(t)-O(t) \Vert \leq \frac{\vlr |t|}{l}  g(l) |A| \Vert O \Vert,
\ee
and also,
for any operator $U$ whose support does not intersect $b_l(A)$, we have
\be
\Vert [O(t),U] \Vert \leq \frac{\vlr |t|}{l}  g(l)|A|  \Vert O \Vert \Vert U \Vert,
\ee
where $|A|$ denotes the cardinality of the set $A$ and $g(l)$ decays faster than exponentially in $l/R$.  The quantity
$\vlr$ depends on $R,J$, while the function $g$ depends only on $R$.
\begin{proof}
This is a minor variant of Lieb-Robinson bounds proven previously.  See the appendix for an example of how such bounds
are proven for a more general class of Hamiltonians.
\end{proof}
\end{lemma}
For short times, having the factor of $|t|$ in the above bound is useful, as it will help us deal with
cases later that would otherwise lead to divergent integrals at short times.

\begin{definition}
Given any operator $O$ and function $G$, we define $W_{\gamma,G}(O)$ to be the {\bf operator $O$ filtered below energy $\gamma$} by
\be
W_{\gamma,G}(O)=\gamma \int {\rm d}t G(\gamma t) \exp(i H_0 t) O \exp(-i H_0 t),
\ee
where the filter function $G(t)$ and its corresponding Fourier transform
$\tilde G(\omega)$ are chosen to have the properties that $\tilde G(\omega)$ is close to $0$ for $|\omega|\geq 1$,
$\tilde G(\omega)$ is close to $1$ for $|\omega| \leq 1/2$, and
$G(t)$ is an even function of $t$ and decays rapidly in $t$.
\end{definition}

Two specific examples of such filter functions are the following.  First, we can define
\be
G(t)=f_q(t)=\exp(-t^2/2q) [\exp(i3t/4)-\exp(-i3t/4)]/it.
\ee
For $q\rightarrow \infty$, the Fourier transform of this is a filter onto frequencies
between $-3/4$ and $+3/4$.  That is, it is equal to unity for $|\omega|<3/4$ and equal to zero for $|\omega|>3/4$.
For finite $q$, 
$f_q(t)$
and its corresponding Fourier transform $\tilde f_q(\omega)$ obey the following properties:
\begin{itemize}
\item $|\tilde f_q(\omega)| \leq \exp(-Cq)$ if $|\omega|\geq 1$, for some numeric
constant $C$.

\item $|\tilde f_q(\omega)-1| \leq \exp(-Cq)$ if $|\omega|\leq 1/2$, for some numeric
constant $C$.

\item $|f_q(t)| \leq C \exp(-t^2/2q)$, for some numeric constant $C$.
\end{itemize}
This is an example of a Gaussian filter.
Second, we can define an exact filer.
An exact  filter is a function $G(\omega)=F_{low}(t)$, where $F_{low}(t)$ is an even function, decaying
faster than any power of $t$, with
$\tilde F_{low}(\omega)=1$ for $|\omega|\leq 1/2$ and $\tilde F_{low}(\omega)=0$ for $|\omega|=1$.

If $G(t)$ decays rapidly as a function of $t$, the operator $W_{\gamma,G}(O)$ has the following localizability property which follows from
a Lieb-Robinson bound:
\begin{lemma}
If $O$ is supported on set $A$, then for any $l$, the operator $W_{\gamma,G}(O)$ can be approximated by an operator $W_{\gamma,G}^l(O)$ which is
supported on $b_l(A)$ up to an error
\be
\label{err}
\Vert W_{\gamma,G}(O) - W_{\gamma,G}^l(O) \Vert \leq \Bigl\{\int_{|u|\geq l\gamma/\vlr} |G(u)| {\rm d}u+g(l) |A| \int{\rm d}u|G(u)|
\Bigr\} \Vert O \Vert.
\ee
\begin{proof}
Let $O'$ be defined by
\be
O'=
\gamma \int_{-l/\vlr}^{l/\vlr} {\rm d}t G(\gamma t) \exp(i H_0 t) O \exp(-i H_0 t).
\ee
By a triangle inequality,
\begin{eqnarray}
\label{eq1}
\Vert  O'-O \Vert  & \leq & \gamma \int_{|t|\geq l/\vlr} 
{\rm d}t 
|G(\gamma t)| 
\Vert O \Vert
\\ \nonumber
&=& 
\int_{|u|\geq l\gamma/\vlr} 
{\rm d}u 
|G(u)| 
\Vert O \Vert 
\end{eqnarray}

Then, define $O_l$ by
\be
O_l=
\gamma \int_{-l/\vlr}^{l/\vlr} {\rm d}t G(\gamma t) O_l(t).
\ee
By a triangle inequality and the Lieb-Robinson bound,
\begin{eqnarray}
\label{eq2}
\Vert O_l-O' \Vert &\leq &
\gamma \int_{-l/\vlr}^{l/\vlr} 
{\rm d}t 
|G(\gamma t)| 
g(l) |A| \Vert O \Vert \\ \nonumber
&\leq & \gamma \int {\rm d}t |G(\gamma t)| g(l) |A| \Vert O \Vert \\ \nonumber
&\leq & \int {\rm d}u |G(u)| g(l) |A| \Vert O \Vert.
\end{eqnarray}
Eq.~(\ref{err}) follows from Eqs.~(\ref{eq1},\ref{eq2}) by a triangle inequality.
\end{proof}
\end{lemma}

As a corollary of the above result, we find that
\begin{corollary}
If $O$ is supported on set $A$, then for any $l$, the operator $W_{\gamma,f_q}(O)$ can be approximated by an operator $W_{\gamma,f_q}^l(O)$ which is
supported on $b_l(A)$ up to an error
\be
\Vert W_{\gamma,f_q}(O) - W_{\gamma,f_q}^l(O) \Vert \leq \Bigl( C \frac{q}{l\gamma/\vlr}\exp[-(l \gamma/\vlr)^2/2q]+C \sqrt{q} g(l) |A| \Bigr) \Vert O \Vert.
\ee
\end{corollary}
and also
\begin{corollary}
If $O$ is supported on set $A$, then for any $l$, the operator $W_{\gamma,F_{low}}(O)$ can be approximated by an operator $W_{\gamma,F_{low}}^l(O)$ which is
supported on $b_l(A)$ up to an error bounded by
$|A| \Vert O \Vert$ times a function decaying faster than any power of $l$.
\end{corollary}

\begin{definition}
A Hamiltonian $H$ is said to have a {\bf mobility gap} $\gamma$ and localization length $\xi$ and localization constant $\con$ up to
time $t_{max}$ if,
for any operator $O$ supported on set $A$ and any filter function $G$, and any $t$ with $|t|\leq t_{max}$,
there exists an operator
$W_{\gamma,G}^{loc}(O,t)$ with the following properties.
First, for any $l$, 
$W_{\gamma,G}^{loc}(O,t)$ can be approximated by an operator supported on $b_l(A)$ up to an error in operator
norm bounded by
\be
\label{leak}
\con\exp(-l/\xi) \Vert W_{\gamma,G}(O) \Vert
+{\rm max}_{|\omega|\geq \gamma} |\tilde G(\omega)|
\Vert O \Vert.
\ee
Second, we
require that the state produced by acting with $W_{\gamma,G}(O)(t)=\exp(i H_0 t) W_{\gamma,G}(O)\exp(-i H_0 t)$ on the ground state
is equal to the state produced by acting with the operator $W_{\gamma,G}^{loc}(O,t)$ on the ground state, i.e.,
\be
W_{\gamma,q}(O)(t) \Psi_0 = W_{\gamma,G}^{loc}(O,t)  \Psi_0.
\ee
Third, we have
\be
\Vert W_{\gamma,G}^{loc}(O,t) \Vert \leq \Vert W_{\gamma,G}(O) \Vert.
\ee
\end{definition}

The above definition is our many-body version of the single particle definition of localization.  It is an analogue of the
definition of {\it dynamical localization}\cite{dloc}.
The second term on the right-hand side of Eq.~(\ref{leak}) reflects the ``leakage" of states above the
mobility gap due to the approximate nature of the filtering.

It may happen that for a given system, there are several different choices of $\gamma,\xi,\con,t_{max}$ for
which the system has a mobility gap.  For example, in \cite{tjo}, a very strong form of many-body localization was shown:
for any operator $O$ (even without filtering), the operator $\exp(i H_0 t) O \exp(-i H_0 t)$
could be approximated by an operator on a distance $l$ with small error for $l$ that was only logarithmically large in the time.
Hence, by taking a given $\xi$, for the system in \cite{tjo}, one can find a $t_{max}$ that is exponentially large in $\xi$.

We assume that the ground state of $H_0$ has energy $E_0=0$.
We define $\lambda_{min}$ to be the second smallest eigenvalue of $H$.  
We will assume later only very modest requirements on $\lambda_{min}$.  We will need
$t_{max} \lambda_{min}$ to be large, to control errors in quasi-adiabatic continuation.  We will also need a
unique bulk state as defined below: to have this, it suffices, but is not necessary to have
$\lambda_{min}\geq 1/{\rm poly}(L)$.  This is a very
weak requirement; for example, in a single particle system, the eigenvalue distribution is smooth, and so $\lambda_{min}$ will
typically be of order $1/V$.

We now define a corrected quasi-adiabatic continuation operator using the above
definitions.
\begin{definition}
Given a parameter-dependent Hamiltonian, $H_0$, an operator $O$, and functions $F(t),G$ 
we define
the {\bf corrected quasi-adiabatic continuation operator} at mobility scale $\gamma$ and low energy cutoff $\lambda^<$ to  be the operator
${\cal D}(H_0,O)$ defined by
\begin{eqnarray}
\label{cordef}
i{\cal D}(H_0,O)&=&\int 
F(2\gamma t) \exp(i H_0 t) (O-W_{\gamma,G}(O)) \exp(-i H_0 t)
{\rm d}t 
\\ \nonumber
&&+
\int 
F(\lambda^< t) W_{\gamma,G}^{loc}(O,t)
{\rm d}t,
\end{eqnarray}
where the function $F(t)$ has the property that its Fourier transform $\tilde F(\omega)$ obeys
\be
\label{Berryprop}
\tilde F(0)=0,
\ee
and where $F$ is an odd function of time so that ${\cal D}$ is Hermitian.
\end{definition}

Given a parameter dependent Hamiltonian $H_s=\sum_Z H_Z(s)$,
we define
\be
{\cal D}_s={\cal D}(H_s,\partial_s H_s).
\ee
We also sometimes write ${\cal D}^Z_s={\cal D}(H_s,\partial_s H_Z(s))$, so that
\be
\label{de}
{\cal D}_s=\sum_Z {\cal D}^Z_s.
\ee

\begin{definition}
For each such quasi-adiabatic continuation, we define a function
${\cal C}(\omega)$ by
\be
{\cal C}(\omega)
\equiv
\frac{1}{\lambda^<} \tilde F(\omega/\lambda^<) \tilde G(\omega)+ \frac{1}{2 \gamma} \tilde F(\omega/2 \gamma) (1-\tilde G(\omega)).
\ee
\end{definition}
The operator ${\cal D}$ will be used to approximate adiabatic evolution in a local way.  To do this, we will require that
${\cal C}(\omega)$ be close to $-1/\omega$ for $|\omega|\geq \lambda^>$.   See lemma (\ref{lemgausserr}) and lemma (\ref{lemexerr}) where
we will show that
\be
\Bigl| \partial_0 | \Psi_0(s) \rangle - i {\cal D}_{\alpha}(H_0,\partial_s H_s) |\Psi_0 \rangle \Bigr| \leq \Vert \partial_s H_s \Vert {\rm max}_{\omega} |{\cal C}(\omega)+1/\omega|.
\ee

The corrected quasi-adiabatic continuation operator here differs from previous ones, such as Eq.~(\ref{uncorr}),
by the addition of the terms involving
$W_{\gamma,q}^{loc}(\partial_s H_s)$.  This will be used to account for low frequency components below the mobility gap.  Our idea
is as follows: for frequencies above the mobility gap, we use the large frequency to enable us to approximate the adiabatic evolution
by an integral over a short-range of times, and hence with a local operator, while for frequencies below the mobility gap, we
greatly increase the time scale we use to approximate adiabatic evolution, but we use the assumption of localization below the mobility
gap to keep the operators local.

One example of such a function $F$ gives a Gaussian
corrected quasi-adiabatic continuation operator.
In \cite{hall}, the quasi-adiabatic evolution operator was defined by
\be
\label{gauss}
\frac{i}{\alpha\sqrt{2\pi}} \int {\rm d}u \int_0^u {\rm d}t \exp(i H_0 t) O \exp(-i H_0 t) \exp(-u^2/2\alpha^2),
\ee
for some parameter $\alpha$,
while in \cite{lsm,hwen} a more complicated integral was used.
Eq.~(\ref{gauss}) can be re-written as
\be
\int {\rm d}t F(t) \exp(i H_0 t) O \exp(-i H_0 t),
\ee
where $F(t)$ is defined to be 
\be
\frac{i}{\alpha\sqrt{2\pi}}\int_t^{\infty} {\rm d}u \exp(-u^2/2\alpha^2),
\ee
for $t>0$ and $F(t)$ is an odd function.
We can use this function $F(t)$ in our corrected quasi-adiabatic continuation definition, using $G=f_q$ for the filter function, getting
\be
\label{qadERR}
|{\cal C}(\omega)+1/\omega | \leq
C (1/\lambda_{min}) \exp[-C (\lambda^</\lambda_{min})^2\alpha^2]+C(1/\lambda_{min}) \exp[-Cq]+
C(1/\gamma)\exp[-C\alpha^2/2],
\ee
for $|\omega| \geq \lambda_{min}$,
and we have
\be
|F(t)| \leq C \exp[-t^2/2\alpha^2].
\ee

Alternatively, we can define an exact
corrected quasi-adiabatic continuation operator 
at mobility scale $\gamma$, and {\bf low-energy cutoff} $\lambda^<$ by a function $F(t)$
where
$F(t)$ is some function which has the property that its Fourier transform, $\tilde F(\omega)$, is odd and infinitely differentiable and equals
$-1/\omega$ for $|\omega|\geq 1$.
Then, we have
\be
\label{qadERRsmooth}
{\cal C}(\omega)=-1/\omega 
\ee
for $|\omega| \geq \lambda^<$.
Since $\tilde F(\omega)$ is infinitely differentiable, $F(t)$ decays faster than any power of $t$.

Finally, we need one more definition\cite{ak}:
\begin{definition}
\label{unique}
We say that a Hamiltonian $H$ has an $(l,\gub)$ {\bf unique ground state} if the following holds for all $\epsilon\geq 0$.
Given any density matrix
$\rho$ such that, for all sets $A$ with ${\rm diam}(A)\leq l$ the inequality
\be
\label{rhoclose}
\Vert {\rm Tr}_{\overline A}(\rho-P_0) \Vert_1 \leq \epsilon
\ee
holds, then
\be
\label{cdef}
\Vert \rho-P_0 \Vert_1 \leq \gub \sqrt{\epsilon}.
\ee
\end{definition}

We now show that give a bound on the smallest eigenvalue, $\lambda_{min}$, then an $(l,\gub)$ unique ground state follows for a $\gub$ that depends on $\lambda_{min}$.
\begin{lemma}
\label{spectoloc}
If the second smallest eigenvalue is at least $\lambda_{min}$ for a Hamiltonian $H$ with range $R$ and interaction strength $J$, then it has an
$(R,2 \sqrt{JV/\lambda_{min}})$ unique ground state according to the above definition.
\begin{proof}
Suppose Eq.~(\ref{rhoclose}) holds.
Note that ${\rm Tr}(\rho H)=\sum_Z {\rm Tr}(\rho H_Z)$.  Then, since each $H_Z$ is supported on a set of diameter $R$,
we have that ${\rm Tr}((\rho-P_0) H_Z )\leq \epsilon \Vert H_Z \Vert$.  Summing over $Z$, ${\rm Tr}(\rho H)-E_0) \leq \Vert H \Vert \leq
\epsilon JV$.  We now maximize the 1-norm difference between $\rho$ and $P_0$ subject to this constraint on the
energy difference.  The maximum is obtained when $\rho=|\psi\rangle\langle \psi|$, for
$\psi=\cos(\theta)\Psi_0 + \sin(\theta) \Psi_1$, with $\Psi_1$ being an
eigenstate of $H$ with energy $\lambda_{min}$.  Using the estimate of ${\rm tr}(\rho H)$, $\sin(\theta)^2\leq \epsilon JV/\lambda_{min}$.
In this two dimensional subspace, $\rho-P_0$ equals
\be
\begin{pmatrix}
-\sin(\theta)^2 & \cos(\theta)\sin(\theta) \\
\cos(\theta)\sin(\theta) & \sin(\theta)^2
\end{pmatrix},
\ee
and
$\Vert \rho-P_0 \Vert_1 = 2\sqrt{\sin(\theta)^4+\cos(\theta)^2\sin(\theta)^2}=2 \sin(\theta)
\leq 2\sqrt{\epsilon JV/\lambda_{min}}$.

Thus, Eq.~(\ref{cdef}) follows with $\gub=2\sqrt{JV/\lambda_{min}}$.
\end{proof}
\end{lemma}

One fundamental idea in
\cite{lsm} and \cite{hall}, was to show that quasi-adiabatic evolution around certain closed paths in parameter space left the energy
almost unchanged at the end of the path; then, using the existence of a spectral gap, the fact that the energy
was almost unchanged was used to show that we had returned to a state close to the ground state.
Here, we will use this $(l,\gub)$ unique bulk state assumption instead of a spectral gap since we may not have a spectral gap.  In fact,
however, since we will only need the $(l,\gub)$ assumption for $\gub \geq 1/{\rm poly}(L)$, from lemma (\ref{spectoloc}) it suffices
to have a minimum  eigenvalue $\lambda_{min}$ which is greater than or equal to $1/{\rm poly}(L)$.  So, the $(l,\gub)$ unique
ground state assumption follows from a very weak assumption on $\lambda_{min}$, as claimed above.

This unique ground state assumption is physically necessary
when we prove Hall conductance quantization later
under a weaker assumption of a mobility gap, compared to the spectral gap assumption of \cite{hall}.
On physical grounds, we need to have the $(l,\gub)$ unique ground state for the following reason: consider a fractional
Hall system on a torus with multiply degenerate ground state and then a spectral gap (not just a mobility gap) to the rest
of the spectrum.  This system will {\it not} display integer Hall conductance quantization.  However, it will actually show a mobility
gap, up to exponentially large times $t$, since after filtering any operator acting on one ground state can only have matrix
elements to one of the other ground states, and since the splitting between the ground states is exponentially small in system size,
the filtered operator will be almost unchanging in time.  Thus, we do not expect that there is any Hall conductance quantization
theorem in the absence of some condition like the $(l,\gub)$ unique ground state condition.

As further justification for our $(l,\gub)$ unique ground state definition, we
note that if this definition does not hold for $l=L/2-1$ and $\gub<<1$, then
there exists another state $\psi$ orthogonal to $\Psi_0$ with the property that, given any local operator $O$ with support on a set of diameter less than
half the system size, if the operator $O$ is projected into the two-dimensional
spanned by $\Psi_0,\psi$ it is close to the identity operator.  This is a
definition of topological order (see, for example the definition of $(l,\epsilon)$ topological order in \cite{bhv}).  Thus, since we will only use this
unique ground state definition for $l\sim {\rm const}. \times L$ later, we are in fact only
requiring the absence of topological order.

A further reason for introducing the unique ground state assumption is that later in the context of the Hall effect with boundaries
we will need a different unique bulk state assumption, definition (\ref{uniquebulk}), which generalizes this unique ground state assumption.

\section{Correlation Decay}
The most basic result to show using these definitions is the exponential decay of
correlations in a system with an unique ground state and a mobility gap.
We do all these calculations with Gaussian filter functions.  This section can be read separately from the later sections of the text, because
it only relies on the assumption of a mobility gap and does not use the definitions of quasi-adiabatic continuation operators.

We begin with a lemma:
\begin{lemma}
Let $P_{\gamma/2}$ denote the projector onto eigenstates with energy greater than or equal to
$E_0+\gamma/2$.
Let $O_A,O_B$ be operators supported on sets $A,B$ with
${\rm dist}(A,B)=l$.
Suppose that $O_B$ has that property that
\be
\label{nolow}
\Bigl| (1-P_{\gamma/2}) O_B\Psi_0 \Bigr| \leq \delta,
\ee
for some $\delta$.
Then,
\be
|\langle \Psi_0, O_A O_B \Psi_0 \rangle -
\langle \Psi_0, O_A \Psi_0 \rangle \langle \Psi_0, O_B \Psi_0 \rangle|
\leq
C \Bigl\{ \exp(-Cl\gamma/2 \vlr)+{\rm min}(|A|,|B|) g(l) +\delta\Bigr\} \Vert O_A \Vert \Vert O_B \Vert.
\ee.
\begin{proof}
The proof basically follows previously proven correlation bounds\cite{lr2}.
Assume without loss of generality that $\langle \Psi_0,O_A \Psi_0 \rangle =\langle \Psi_0, O_B \Psi_0 \rangle=0$.
For any operator $X$, following \cite{net}, we define $\tilde X^+$ by
\be
\label{tdef}
\tilde X^+=\lim_{\epsilon\rightarrow 0^+} \frac{1}{2\pi}\int {\rm d}t \exp(i H_0 t) X \exp(-i H_0 t)
\frac{\exp[-(t\gamma/2)^2/2q]} {it+\epsilon},
\ee
for some $q$ which will be chosen equal to $l\gamma/2\vlr$ below.

The operator $\tilde X^+$ is equal to $X(t)=\exp(i H_0 t)$ convolved against the function
$(1/2\pi) (\exp[-(t\gamma/2)^2/2q])/(it+\epsilon)$.  For $q\rightarrow \infty$, the Fourier function converges to a step function,
vanishing for negative $\omega$, and unity for positive $\omega$.
For finite $q$, one may show that for any such operator $X$, we have
\be
\Bigl| P_{\gamma/2}  \Bigl(X- \tilde X^+\Bigr) |\Psi_0\rangle \Bigr| \leq C \frac{\exp(-q/2)}{\sqrt{2\pi q}} \Vert X \Vert,
\ee
as shown in \cite{net}, and also that
\be
\Bigl| \langle \Psi_0| \Bigl( X-\tilde X^+\Bigr)P_{\gamma/2}
\Bigr| \leq C \frac{\exp(-q/2)}{\sqrt{2\pi q}} \Vert X \Vert.
\ee
Further, one may show from Eq.~(\ref{nolow}) that
\be
\Bigl| (1-P_{\gamma/2}) \Bigl( O_B|\Psi_0\rangle-\tilde O_B^+ |\Psi_0\rangle\Bigr) \Bigr| \leq \delta.
\ee

Thus,
\be
\Bigl| \Bigl(\tilde O_B^+-O_B\Bigr) | \Psi_0\rangle \Bigr|\leq 
C \frac{\exp(-q/2)}{\sqrt{2\pi q}} \Vert O_B \Vert+ \delta.
\ee

We now estimate the commutator
$[\tilde O_B^+,O_A]$ by the Lieb-Robinson bound and the usual trick of 
splitting the time integral into integrals over early times ($|t|\leq l/\vlr$) and the integral over late times ($|t|\geq l/\vlr$).
By a triangle inequality
\begin{eqnarray}
\Vert [ \tilde O_B^+,O_A] \Vert & \leq &
\frac{1}{2\pi} \int {\rm d}t 
\frac{\exp[-(t\gamma/2)^2/2q]} {|t|}
\Vert [O_B(t),O_A] \Vert
\\ \nonumber
&\leq &\frac{1}{2\pi} \int_{|t|\geq l/\vlr} {\rm d}t 
\frac{\exp[-(t\gamma/2)^2/2q]} {|t|}
\Vert [O_B(t),O_A] \Vert+
\frac{1}{2\pi}\int_{|t|<l/\vlr} {\rm d}t \frac{\vlr}{l} |B| g(l)
\Vert [O_B(t),O_A] \Vert \\ \nonumber
&\leq &\frac{2}{2\pi} \int_{|t|\geq l/\vlr} {\rm d}t 
\frac{\exp[-(t\gamma/2)^2/2q]} {|t|}
\Vert O_B(t) \Vert \Vert O_A \Vert+
\frac{1}{2\pi}\int_{|t|<l/\vlr} {\rm d}t \frac{\vlr}{l} |B| g(l)
\Vert [O_B(t),O_A] \Vert \\ \nonumber
&\leq &
C\Bigl\{(q\vlr^2/(l^2\gamma^2)) \exp[-(l^2\gamma^2/4\vlr^2)/2q]+|B| g(l) \Bigr\} \Vert O_A \Vert
\Vert O_B \Vert.
\end{eqnarray}

Using the triangle inequality $|\langle \Psi_0, O_A O_B \Psi_0 \rangle|\leq
|\langle \Psi_0,O_A (O_B-\tilde O_B^+) \Psi_0 \rangle| +
|\langle \Psi_0,\tilde O_B^+ O_A \Psi_0 \rangle|+
\Vert \tilde [O_A,O_B^+] \Vert$,
choosing $q \sim l\gamma/2\vlr$, we find that
\be
|\langle \Psi_0, O_A O_B \Psi_0 \rangle| \leq C \Bigl\{ \frac{\vlr}{l\gamma} \exp(-Cl\gamma/2 \vlr)+g(l)|B| +\delta\Bigr\} \Vert O_A \Vert \Vert O_B \Vert.
\ee 
Since $(\vlr/l\gamma)\exp(-Cl\gamma/2\vlr)\leq C\exp(-l\gamma/2\vlr)$ for $l\geq \vlr/\gamma$, we have
\be
|\langle \Psi_0, O_A O_B \Psi_0 \rangle| \leq C \Bigl\{ \exp(-Cl\gamma/2 \vlr)+g(l)|B| +\delta\Bigr\} \Vert O_A \Vert \Vert O_B \Vert.
\ee
One can replace the $|B|$ with $|A|$ by applying the Lieb-Robinson bound to the
evolution of $O_A$ rather than $O_B$.

\end{proof}
\end{lemma}

\begin{theorem}
Assume that $H$ has a mobility gap $\gamma$,
localization length $\xi$ and localization constant $\con$ up to
time $t_{max}$.  
Let $O_A,O_B$ be operators supported on sets $A,B$ with
${\rm dist}(A,B)=l$.  Assume that the lowest eigenvalue of $H$ is $\lambda_{min}$.
Then,
\be
|\langle \Psi_0 O_A O_B \Psi_0 \rangle| \leq
\Bigl\{ Cg(l/2) |A| + \exp(-t_{max}\lambda_{min})+
\log(2\vlr t_{max}/l) \Bigl( \con\exp(-l/\xi)+C\exp(-Cl\gamma/2\vlr) 
\Bigr)\Bigr\} \Vert O_A \Vert \Vert O_B \Vert.
\ee
\begin{proof}
Assume, without loss of generality, that $\langle O_A \rangle=\langle O_B \rangle=0$.
Define $W_{\gamma,f_q}(O_B)$ to be the operator $O_B$ filtered below energy $\gamma$ as before. 
Define
$Z=W_{\gamma,f_q}^{l/2}(O)$, with $q=l\gamma/2\vlr$.
Then,
\be
\label{splitted}
\langle \Psi_0, O_A O_B \Psi_0 \rangle =\langle \Psi_0, O_A \Bigl( O_B-Z \Bigr) \Psi_0 \rangle
+ \langle \Psi_0, O_A \Bigl( Z-W_{\gamma,f_q}(O)\Bigr) \Psi_0 \rangle
+ \langle \Psi_0, O_A W_{\gamma,f_q}(O) \Psi_0 \rangle.
\ee

Let $X=O_B-Z$.
We have
\begin{eqnarray}
\Bigl| (1-P_{\gamma/2})| X \Psi_0 \rangle \Bigr| & \leq& \Vert 
W_{\gamma,f_q}^{l/2}(O_B)-
W_{\gamma,f_q}(O_B) \Vert + \exp(-Cq) \Vert O_B \Vert
\\ \nonumber
&\leq &
\Bigl( C \frac{q}{l\gamma/2\vlr}\exp[-(l \gamma/2\vlr)^2/2q]+C g(l/2) |A| +\exp(-Cq)\Bigr) \Vert O_B \Vert.
\end{eqnarray}
Further, $X$ is supported on a set which is at least distance $l/2$ from $A$.  So,
by the previous lemma,
\begin{eqnarray}
|\langle \Psi_0,O_A X \Psi_0 \rangle| & \leq &
C \Bigl\{ \exp(-Cl\gamma/2 \vlr)+|A| g(l) +
C g(l/2) |A| +\exp(-Cq)\Bigr) 
\Bigr\} \Vert O_A \Vert \Vert O_B \Vert
\\ \nonumber
&\leq &
C \Bigl\{ \exp(-Cq)+|A| g(l/2)\Bigr)
\Bigr\} \Vert O_A \Vert \Vert O_B \Vert.
\end{eqnarray}
Note that if desired, the term $|A|$ in the above expression can be replaced by
the cardinality of the set of sites within distance $l/2$ of $B$ and the bound still holds.

We now consider the term
$\langle \Psi_0, O_A \Bigl( Z-W_{\gamma,f_q}(O)\Bigr) \Psi_0 \rangle$ 
in Eq.~(\ref{splitted}).
This is bounded by
\begin{eqnarray}
&& \Vert W_{\gamma,f_q}^{l/2}(O_B)-
W_{\gamma,f_q}(O_B) \Vert
\\ \nonumber
&\leq &
\Bigl( C \frac{q}{l\gamma/2\vlr}\exp[-(l \gamma/2\vlr)^2/2q]+C g(l/2) |A|\Bigr) \Vert O_B \Vert.
\end{eqnarray}

We finally consider the term
$\langle \Psi_0, O_A W_{\gamma,f_q}(O) \Psi_0 \rangle$. 
Define $Y$ by
\begin{eqnarray}
\label{Ydef}
Y&=&
\int {\rm d}t_{|t|\leq t_{max}} W_{\gamma,f_q}^{loc}(O,t) \fl(t)
\\ \nonumber
&&+\int {\rm d}t_{|t|\geq t_{max}} \exp(i H_0 t) W_{\gamma,f_q}(O) \exp(-i H_0 t) \fl(t),
\end{eqnarray}
where the function $\fl(t)$ is defined by
\be
\fl(t)=\exp[-(t\lambda_{min})^2/2q_1] \frac{\exp(i 2 \gamma t)-1}{i t},
\ee
for some $q_1$.
This is the similar to (\ref{tdef}), but we use the filter $\fl(t)$.  For $q_1\rightarrow\infty$, the filter $\fl(t)$ has a Fourier
transform equal to unity for $0<\omega<2\gamma t$ and vanishing for $\omega<0$ or $\omega>2\gamma t$ and so at infinite
$q_1$, $|W_{\gamma,f_q}(O)\Psi_0-Y\Psi_0|\leq |P_{2\gamma}W_{\gamma,f_q}(O) \Psi_0| \leq \exp(-C q) \Vert O \Vert$.
For finite $q_1$, the
filter $\fl(t)$ approximates this filter.  The filter $\fl(t)$ is chosen not to have a singularity at $t=0$.
So, using the assumption on $\lambda_{min}$, we find that
\be
\Bigl| W_{\gamma,f_q}(O) |\Psi_0\rangle-Y |\Psi_0\rangle \Bigr| \leq
(\exp(-Cq)+\frac{\exp(-q_1/2)}{\sqrt{2\pi q_1}}) \Vert O_B \Vert.
\ee
Similarly,
\be
\Bigl| \langle \Psi_0| Y \Bigr| \leq
(\exp(-Cq)+\frac{\exp(-q_1/2)}{\sqrt{2\pi q_1}} \Vert O_B \Vert.
\ee

We now estimate the commutator
$\Vert [Y,O_A] \Vert$.  We do this using a triangle inequality.
The integral over $|t|\geq t_{max}$ in (\ref{Ydef}) is bounded by $C (q_1/t_{max}\lambda_{min}) \exp[-(t_{max}\lambda_{min})^2/2q_1]$.
We break the integral up of $|t|\leq t_{max}$ into two different parts.  First, a part with $|t|\leq l/2\vlr$.
Second, a part with $l/2\vlr \leq |t| \leq t_{max}$.

We now bound the integral over $|t| \leq l/2\vlr$.
By the localization assumption,
\be
\Vert [W_{\gamma,f_q}(O_B,t),O_A]\Vert \leq \Bigl( \con\exp(-l/\xi)+C\exp(-Cq) \Bigr) \Vert O_A
\Vert \Vert O_B \Vert.
\ee
For 
$|t| \leq l/2\vlr$, $\fl(t)$ is bounded by $2\gamma$.  
Thus, the integral over
$|t| \leq l/2\vlr$ is bounded by
\be
(\gamma l/\vlr) \Bigl( \con\exp(-l/\xi)+C\exp(-cq) \Bigr) \Vert O_A \Vert \Vert O_B \Vert.
\ee

We next bound the integral over $l/2\vlr \leq |t| \leq t_{max}$.
By the localization assumption,
\be
\Vert [W_{\gamma,f_q}(O_B,t),O_A]\Vert \leq \Bigl( \con\exp(-l/\xi)+C\exp(-Cq) \Bigr) \Vert O_A
\Vert \Vert O_B \Vert.
\ee
Thus, the integral over
$l/2\vlr \leq |t| \leq t_{max}$ is bounded by
\be
\log(2\vlr t_{max}/l) \Bigl( \con\exp(-l/\xi)+C\exp(-cq) \Bigr) \Vert O_A \Vert \Vert O_B \Vert.
\ee

We pick $q_1=t_{max}\lambda_{min}$.
Thus by a triangle inequality,
\be
|\langle \Psi_0, O_A Z \Psi_0 \rangle| \leq
\Bigl\{ Cg(l/2) |A| + \exp(-t_{max}\lambda_{min})+
\log(2\vlr t_{max}/l) \Bigl( \con\exp(-l/\xi)+C\exp(-Cl\gamma/2\vlr) 
\Bigr)\Bigr\} \Vert O_A \Vert \Vert O_B \Vert.
\ee
Thus,
\be
|\langle \Psi_0 O_A O_B \Psi_0 \rangle| \leq
\Bigl\{ Cg(l/2) |A| + \exp(-t_{max}\lambda_{min})+
\log(2\vlr t_{max}/l) \Bigl( \con\exp(-l/\xi)+C\exp(-Cl\gamma/2\vlr) 
\Bigr)\Bigr\} \Vert O_A \Vert \Vert O_B \Vert.
\ee
\end{proof}
\end{theorem}

\section{Properties of Corrected Quasi-Adiabatic Continuation Operator}
We now consider the properties of the corrected quasi-adiabatic continuation operator.
There are three basic properties used previously in studying these
systems.
First, the quasi-adiabatic continuation
operator should be local, in that it should be a sum of operators, each of which is
exponentially decaying in space, in the sense that each such operator can be approximated
to exponentially good accuracy by an operator with finite range.
Second, the quasi-adiabatic continuation operator should approximate the exact
adiabatic evolution of a state in a region in which the system has a mobility gap
and a sufficiently large $\lambda_{min}$.  Third, the quasi-adiabatic continuation
operator should produce the correct Berry phase: that is, we should have that
\be
\langle \Psi_0(s),
{\cal D}(H_s,\partial_s H_s) \Psi_0(s) \rangle=0,
\ee
where $\Psi_0(s)$ is the ground state of $H_s$.
This last property follows immediately from Eq.~(\ref{Berryprop}) in the definition of the corrected quasi-adiabatic continuation operator and was emphasized in \cite{berry1}.  We now show the other two properties for Gaussian and exact filter functions.

\begin{lemma}
\label{locallemma}
Consider a corrected quasi-adiabatic continuation operator and a Hamiltonian with a mobility
gap.  Define a function $\B(t)$ to be the convolution of $F(\gamma t)$ with $\delta(t)-\gamma G(\gamma t)$ (here, $\delta(t)$ denotes the Dirac delta-function).
Then, for any operator $O$ with support on set $A$ and any $U$ with support on set $B$ with
${\rm dist}(A,B)\geq l$, then
\be
\label{qaddecay}
\Vert [{\cal D}(H_0,O), U ]\Vert 
\leq  E(l) \Vert O \Vert \Vert U \Vert,
\ee
where
$E(l)$ is defined by
\begin{eqnarray}
\label{Eldef}
E(l)&\equiv &
\Bigl\{ {\rm max}_t(|\B(t)|)
(l/\vlr) |O| g(l) 
+\int_{|t|\geq l/\vlr} {\rm d}t |\B(t)| \Bigr\}
\\ \nonumber
&+&
\Bigl\{
t_{max}
\Bigl(\con\exp(-l/\xi) 
+{\rm max}_{|\omega|\geq \gamma} |\tilde G(\omega)|\Bigr)
+\int_{|t|\geq t_{max}} {\rm d}t |F(\lambda^< t)| \Bigr\}. 
\end{eqnarray}
and further, for any $l$, there exists an operator $O'$ with support on $b_l(A)$ such that
\begin{eqnarray}
\label{localizeit}
&& \Vert {\cal D}(H_0,O)- O'\Vert  \\ \nonumber
&\leq & E(l) \Vert O \Vert.
\end{eqnarray}
\begin{proof}
We consider both terms in the definition of ${\cal D}$, Eq.~(\ref{cordef}), separately.
For the first term, we wish to bound
\be
\Vert \int {\rm d}t F(t) \exp(i H_0 t) (O-W_{\gamma,G}(O)) \exp(-i H_0 t) ,U] \Vert.
\ee
We note that the operator
\be
 \int {\rm d}t F(t) \exp(i H_0 t) (O-W_{\gamma,G}(O)) \exp(-i H_0 t)
\ee
is equal to
\be
\int {\rm d}t \B(t) \exp(i H_0 t)O \exp(-i H_0 t),
\ee
for a function $\B(t)$ equal to the convolution of $F(t)$ with $\delta(t)-\gamma G(\gamma t)$.

We use a triangle inequality:
\begin{eqnarray}
&& \Vert [\int {\rm d}t \B(t) \exp(i H_0 t) O \exp(-i H_0 t),U ] \Vert
\\ \nonumber
& \leq &
\int_{|t|\leq l/\vlr} {\rm d}t \B(t) \Vert [\exp(i H_0 t) O \exp(-i H_0 t),U ] \Vert  \\ \nonumber
&&+
\int_{|t|\geq l/\vlr} {\rm d}t \B(t) \Vert [\exp(i H_0 t) O \exp(-i H_0 t),U ] \Vert.
\end{eqnarray}
The first term is bounded using the Lieb-Robinson bound by
\be
\label{decay1}
{\rm max}_t(|\B(t)|)
(l/\vlr) |O| g(l) \Vert O \Vert \Vert U \Vert.
\ee
The second term is bounded by
\be
\label{decay2}
\int_{|t|\geq l/\vlr} {\rm d}t |\B(t)|
 \Vert O \Vert \Vert U \Vert.
\ee

We now consider the term
\begin{eqnarray}
&& 
\Vert[
\int {\rm d}t F(\lambda^< t) W_{\gamma,G}^{loc}(O,t),U
]\Vert \\ \nonumber
&\leq &
\int_{|t|\leq t_{max}} {\rm d}t F(\lambda^< t) \Vert [W_{\gamma,G}^{loc}(O,t),U]\Vert \\ \nonumber
&&+\int_{|t|\geq t_{max}} {\rm d}t F(\lambda^< t) \Vert [W_{\gamma,G}^{loc}(O,t),U]\Vert.
\end{eqnarray}

To bound the integral over $|t|\leq t_{max}$, we use the localization assumption, to bound this
by
\be
\label{decay3}
t_{max}
\Bigl(\con\exp(-l/\xi) 
+{\rm max}_{|\omega|\geq \gamma} |\tilde G(\omega)|\Bigr)
\Vert O \Vert  \Vert U \Vert.
\ee

The integral over $|t|\geq t_{max}$ is bounded by
\be
\label{decay4}
\int_{|t|\geq t_{max}} {\rm d}t |F(\lambda^< t)|  \Vert O \Vert \Vert U \Vert.
\ee

Putting these results (\ref{decay1},\ref{decay2},\ref{decay3},\ref{decay4})
together, Eq.~(\ref{qaddecay}) follows.  Since Eq.~(\ref{qaddecay}) holds for
all operators $U$, Eq.~(\ref{localizeit}) follows: to see this, 
define $O'$ to be
\be
O'=\int {\rm d}U  \;
U {\cal D}(H_0,O) U^{\dagger},
\ee
where the integral is over all unitary rotations over sites not in $b_l(A)$ with
the Haar measure.  This trick was introduced in \cite{bhv}.
\end{proof}
\end{lemma}

Applying these results to the Gaussian corrected quasi-adiabatic continuation operator and Gaussian filter function, we have the corollary:
\begin{corollary}
For a Gaussian corrected quasi-adiabatic continuation operator and a Hamiltonian with a mobility
gap, the error term
$E(l)$ in (\ref{Eldef}) is bounded by:
\begin{eqnarray}
&& \Vert [{\cal D}(H_0,O), U ]\Vert  \\ \nonumber
&\leq &
\Bigl\{ C (l/\vlr) |O| g(l)+
C
(l\gamma/\vlr) (l/\vlr)
\Bigl( \alpha^2 \exp[-C(l\gamma/\vlr)^2/\alpha^2]+
q \exp[-C(l\gamma/\vlr)^2/q] \Bigr)
\\ \nonumber
&&
+
t_{max}
\Bigl( \con \exp(-l/\xi)+ C\exp(-Cq) \Bigr)
+\frac{Ct_{max} \alpha^2}{(t_{max}\lambda^<)^2}\exp[-(\lambda^< t_{max}/\alpha)^2/2]  \Bigr\}
\\ \nonumber
&& \times
\Vert O \Vert \Vert U \Vert.
\end{eqnarray}
\begin{proof}
Note that $F(t)$ decays as $\exp[-t/\alpha^2]$, while
$G(t)$ decays as
$\exp[-Ct^2/q]$, so $\B(t)$ is bounded by
$(\gamma t)\Bigl(\exp[-C(\gamma t)^2/\alpha^2]+
\exp[-C(\gamma t)^2/q]\Bigr) $
Note that the function $\B(t)$ decays as $\exp[-C (t/\alpha)^2]$ for large $t$ for some $C$.
The rest follows immediately from the definitions.
\end{proof}
\end{corollary}

Similarly,  for exact corrected quasi-adiabatic continuation operators we have the corollary:
\begin{corollary}
For an exact corrected quasi-adiabatic continuation operator and a Hamiltonian with a mobility
gap, the error term
$E(l)$ in (\ref{Eldef}) is bounded by:
\begin{eqnarray}
\label{qaddecayexact}
E(l) &\leq &
C (l/\vlr) |O| g(l)+ \frac{1}{\gamma} Q_1(l\gamma/\vlr)
+\con t_{max} \exp(-l/\xi)
+\frac{1}{\lambda^<}Q_2(\lambda^< t_{max}),
\end{eqnarray}
where the functions $Q_1,Q_2$ decays faster than any power of their arguments.
\end{corollary}

This implies the following ``superpolynomial localizability" property:
\begin{corollary}
The exact corrected quasi-adiabatic continuation operator ${\cal D}(H_s,\partial_s H_Z(s))$ can be
approximated by an operator supported on $b_l(Z)$ up to an error bounded by
$|Z| \Vert \partial_s H_Z(s) \Vert$ times
$1/\gamma$ times a function decaying superpolynomially in $l$ plus $1/\lambda^<$ times a function
decaying superpolynomially in $\lambda^< t_{max}$.
\end{corollary}

We now show that this corrected quasi-adiabatic continuation operator approximates
the adiabatic evolution of states for Hamiltonians with a mobility gap.
\begin{lemma}
\label{lemgausserr}
For a Gaussian corrected quasi-adiabatic continuation operator ${\cal D}$, and a Hamiltonian
$H_0$ with a mobility gap, and a given $\lambda_{min}>0$,
we have
\begin{eqnarray}
\label{qadapprox}
&& \Bigl| \partial_s | \Psi_0(s) \rangle - i {\cal D}_{s}(H_0,\partial_s H_s) |\Psi_0 \rangle \Bigr|  
\\ \nonumber & \leq &
|\tilde F(\omega/\lambda^<) \tilde f(\omega,q)+ \tilde F(\omega/\gamma) (1-\tilde f(\omega,q))+1/\omega| 
\Vert \partial_s H_s \Vert
\\ \nonumber & \leq & 
C \Bigl( (1/\lambda_{min}) \exp[-C (\lambda^</\lambda_{min})^2\alpha^2]+(1/\lambda_{min}) \exp[-Cq]+
(1/\gamma)\exp[-C\alpha^2/2],
\Bigr)
\Vert \partial_s H_s \Vert.
\end{eqnarray}
where $\Psi_0(s)$ is the ground state eigenvector of $H_s$ and the partial derivatives
are taken at $s=0$.
\begin{proof}
Let $\Psi_i(s)$ denote a complete basis of eigenstates of $H_s$, with corresponding
eigenvalues $E_i(s)$.
We have
\be
\partial_s \Psi_0(s)=\sum_{i\neq 0}\frac{1}{E_0(0)-E_i(0)} \Psi_i(s) \langle \Psi_i(0), \Bigl(\partial_s H_s \Bigr) \Psi_0(0) \rangle,
\ee
by linear perturbation theory.

Also, by the definition of the corrected quasi-adiabatic continuation operator,
\be
i{\cal D}_\alpha(H_0,\partial_s H_s) \Psi_0=\sum_{i\neq 0} {\cal C}(E_i-E_0) \Psi_0(0)
\langle \Psi_i(0),\Bigl(\partial_s H_s \Bigr) \Psi_0(0) \rangle,
\ee
so 
\be
\label{whycalC}
\Bigl| \partial_0 | \Psi_0(s) \rangle - i {\cal D}_{\alpha}(H_0,\partial_s H_s) |\Psi_0 \rangle \Bigr| \leq \Vert \partial_s H_s \Vert {\rm max}_{\omega} |{\cal C}(\omega)+1/\omega|.
\ee
so by
Eq.~(\ref{qadERR}), Eq.~(\ref{qadapprox}) follows.
\end{proof}
\end{lemma}

The exact version of the above lemma is much simpler:
\begin{lemma}
\label{lemexerr}
For an exact  corrected quasi-adiabatic continuation operator ${\cal D}$ and exact filter function $F_{low}$, and a Hamiltonian
$H_0$ with a mobility gap, and a given $\lambda_{min} \geq \lambda^<$,
we have
\be
\partial_s | \Psi_0(s) \rangle =i {\cal D}_{s}(H_0,\partial_s H_s) |\Psi_0 \rangle.
\ee
\begin{proof}
This proof is immediate from Eq.~(\ref{qadERRsmooth}).
\end{proof}
\end{lemma}

In some cases, the Gaussian operators above allow tighter estimates.  
However, from now on, for simplicity of estimates, and for the simplicity of
expressing the results, we will use the exact quasi-adiabatic operators (further, using the construction in the appendix, one
can see that in fact our bounds here, which will be expressed only as superpolynomial decay in $L$, in fact become
``subexponential" decay in $L$, as defined in the appendix).  
In particular, the Gaussian operators give slightly better bounds in the Lieb-Schultz-Mattis case of the next
section (we omit the results for simplicity), while the exact operators will actually lead to better bounds in the
Hall conductance section.

\section{Lieb-Schultz-Mattis-type Theorems}
\label{LSM}
We now consider applying these results to prove Lieb-Schultz-Mattis-type theorems.
The ideas here will be needed for the Hall discussion later.  We consider
a system of linear size $L$.  For definiteness, we consider hypercubic geometry; that is,
we consider square geometry in two dimensions ($L$ by $L$), cubic geometry in three
dimensions, and so on.

In this section, we will assume periodic boundary conditions in one direction of the
hypercube, which we call the $\hat x$ direction.  This does not mean that we assume
translation invariance in that direction as done in \cite{lsm1,lsm}.  Rather it means
that our metric ${\rm dist}(\cdot,\cdot)$ only measures distances between points
${\rm mod}(L)$.

The one dimensional Lieb-Schultz-Mattis theorem, as later generalized in \cite{affl}, was a statement about
translationally invariant one-dimensional quantum systems, with finite range and finite strength interaction, and with a conserved local charge.  Having a conserved local
charge means that there is some operator $q_i$, defined on each site $i$, such that
$Q=\sum_i q_i$ commutes with the Hamiltonian, and such that $q_i$ has integer eigenvalues
with $\Vert q_i \Vert \leq q_{max}$ for some given $q_{max}$.  Then, assuming that
$Q$ is not an integer multiple of $L$, it was proven that the gap from the ground
state to the first excited state decays as $1/L$.

This was generalized to higher dimensions in \cite{lsm}.  The most general statement,
in \cite{suffcon}, is that if $Q$ is not an
integer multiple of $L$, then the gap from the ground state to the first excited state
decays is bounded by ${\cal O}(\log(L)/L)$.  

It is important to understand what being an integer multiple of $L$
means.  The work \cite{lsm,suffcon}, only required translational invariance in one direction, the $\hat x$ direction.
However, if the system is an $L$-by-$L'$ torus, and has
a filling fraction $Q/V=1/2$,
then $Q/L$ is non-integer if $L'$ is odd.
The restriction to odd width arises because we use ideas of flux insertion to construct
a state which has low energy and which has a different momentum compared to the ground state,
thus proving bounds on the energy gap variationally.  The major improvement compared to the
one-dimensional result was the ability to handle systems whose aspect ratio was of order unity.

In this section we consider disordered systems, without translation invariance.
Thus, we will certainly {\it not} be able to prove the existence of
low energy excitations in this section, because in the absence of translation invariance there exist Hamiltonians in which $Q$ is not an
integer multiple of $L$ but with a unique ground state and a spectral gap.
Instead, what we will show is the following.  We will construct a flux insertion
operator which inserts $2\pi$ flux in a vertical line, and apply it to the ground state.  We will show that the expectation value of the energy of the resulting state is exponentially small in $L$.
Thus, either $\lambda_{min}$ is exponentially small in $L$, or the flux insertion operator acting on the ground state produces
a state which is superpolynomially close to the ground state multiplied
by a phase.  We then show in the latter case, where the flux insertion operator is superpolynomially close to acting on the ground state by
a phase, that if $Q/L$ is non-integer, then this phase depends in a particular way described below on which line is chosen
for the flux insertion.

To define the flux insertion operator, we need to define the Hamiltonian with twisted
boundary conditions.  Let $Q_X$ be defined by
\be
Q_X =\sum_{i}^{1\leq x(i)\leq L/2} q_i,
\ee
where $x(i)$ is the $\hat x$-coordinate of site $i$.  That is, $Q_X$ is the total charge in
the half of the system to the left of the vertical line with $x=L/2+1$ and to the right of $x=0$.
Let
\be
H(\theta_1,\theta_2)=\sum_Z H_Z(\theta_1,\theta_2),
\ee
where $H_Z(\theta_1,\theta_2)$ is defined as follows.  
If the set $Z$ is within distance $R$ of
the vertical line $x=0$, then $H_Z(\theta_1,\theta_2)=\exp(i \theta_1 Q_X) H_Z \exp(-i \theta_1 Q_X)$;
if the set $Z$ is within distance $R$ of
the vertical line $x=L/2$, then $H_Z(\theta_1,\theta_2)=\exp(-i \theta_2 Q_X) H_Z \exp(i \theta_2 Q_X)$;
otherwise, $H_Z(\theta_1,\theta_2)=H_Z$.
Note that,
\be
\label{equivalent}
H(\theta,-\theta)=\exp(i \theta Q_X) H \exp(-i \theta Q_X).
\ee

We define an operator ${\cal D}_s$ to be an exact corrected quasi-adiabatic continuation
operator with $\lambda^<=\lambda_{min}$ describing quasi-adiabatic continuation along the path from $\theta_1=s$,$\theta_2=-s$.
We have that $\partial_s H(s)=\sum_Z \partial_s H_Z(s)$, and
$\partial_s H_Z(s)$ is nonvanishing if $Z$ is within distance $R$ of the line at $x=0$ or if $Z$ is within
distance $R$ of the line at $x=L/2$.
Let $O^{(1)}(s)$ denote the sum of terms in $\partial_s H(s)$ near the line at $x=0$ and let
$O^{(2)}(s)$ denote the sum of terms in $\partial_s H(s)$ near the line at $x=L/2$, so that
${\cal D}_s={\cal D}(H_s,O^{(1)}(s))+
{\cal D}(H_s,O^{(2)}(s))$.
Note that the Hamiltonians $H_s$ are all unitarily equivalent.  So, if there is a mobility gap at $s=0$, then
there is a mobility gap for all $s$.
Under the assumption of a mobility gap, because of the superpolynomial localizability property,
we can approximate the operators ${\cal D}(H_s,O^{(1)})$ and ${\cal D}(H_s,O^{(2)})$ by operators ${\cal D}_s^{(1)}$ and
${\cal D}_s^{(2)}$ supported within distance less than $L/8$ of the respective lines
$x=0$ and $x=L/2$ up to superpolynomially small error. 
Note that the supports of ${\cal D}_s^{(1,2)}$ do not overlap.

We define the flux insertion operator, $W_1$, as follows.
We define a unitary $U^{(1)}_s$ by
\be
U^{(1)}_s={\cal S}' \exp\{i\int_0^{s} {\rm d}s' {\cal D}^{(1)}_{s'} \},
\ee
where ${\cal S}'$ denotes that the integral is $s'$-ordered.  Then, we set
$W_1=U^{(1)}_{2\pi}$.
We define $W_2$ similarly:
we define a unitary $U^{(2)}_s$ by
\be
U^{(2)}_s={\cal S}' \exp\{i\int_0^{s} {\rm d}s' {\cal D}^{(1)}_{s'} \},
\ee
and we set
$W_2=U^{(2)}_{2\pi}$.

We define
\be
W=W_1 W_2.
\ee

Now, we claim that:
\begin{lemma}
\label{l8}
Assume that the system has an $(L/8,\gub)$ unique bulk state, with $\gub$ greater than or equal to $1/{\rm poly}(L)$.
Assume that the system has a mobility gap, with $t_{max}$ superpolynomially large in $L$.  Then,
\be
{\rm min}_{z_1,|z_1|=1}\Bigl( \Bigl|W_1 |\Psi_0 \rangle- z_1 |\Psi_0\rangle\Bigr| \Bigr)
\ee
is bounded by 
$J/\gamma$ times a function decaying superpolynomially in $L$ plus ${\rm poly}(L) J/\lambda^<$ times a function
decaying superpolynomially in $\lambda^< t_{max}$.
\begin{proof}
Consider any operator $O$ supported on a set of size $A$ at most $L/8$.  If $A$ is not within distance $L/8$ of
the line $x=0$, then
\be
\label{far1}
\langle \Psi_0| W_1^\dagger O W_1 | \Psi_0 \rangle=
\langle \Psi_0| O | \Psi_0 \rangle,
\ee
since $W_1$ commutes with $O$ and $W_1$ is unitary.

On the other hand, if $O$ is within distance $L/8$ of the line $x=0$, then
\begin{eqnarray}
\label{close1}
\langle \Psi_0| W_1^\dagger O W_1 | \Psi_0 \rangle&= &
\langle \Psi_0| W_2^\dagger W_1^\dagger O W_1 W_2| \Psi_0 \rangle \\ \nonumber
&=& \langle \Psi_0| W^\dagger O W| \Psi_0 \rangle,
\end{eqnarray}
since $W_2$ commutes with $W_1$ and $O$.

Define
\be
U={\cal S}' \exp\{ i \int_0^{2\pi} {\rm d}s' {\cal D}_{s'}\},
\ee
along the path $s=\theta_1=-\theta_2$.  Then, the operator norm
difference, $\Vert U-W \Vert$
is bounded by 
$J/\gamma$ times a function decaying superpolynomially in $L$ plus ${\rm poly}(L) J/\lambda^<$ times a function
decaying superpolynomially in $\lambda^< t_{max}$.

However, $U$ is an exact quasi-adiabatic evolution, and the Hamiltonians $H_s$ are unitarily equivalent by Eq.~(\ref{equivalent}).
Thus,
\be
U|\Psi_0\rangle=z |\Psi_0 \rangle,
\ee
for some $z$ with $|z|=1$.
Thus,
$|\langle \Psi_0| W^\dagger O W| \Psi_0 \rangle-\langle \Psi_0| O |\Psi_0 \rangle|$ is bounded by
is bounded by 
$J/\gamma$ times a function decaying superpolynomially in $L$ plus ${\rm poly}(L) J/\lambda^<$ times a function
decaying superpolynomially in $\lambda^< t_{max}$.

Thus, from Eqs.~(\ref{far1},\ref{close1}), for any operator $O$,
\be
|\langle \Psi_0| W_1^\dagger O W_1 | \Psi_0 \rangle-
\langle \Psi_0| O| \Psi_0 \rangle |
\ee
is bounded by 
$J/\gamma$ times a function decaying superpolynomially in $L$ plus ${\rm poly}(L) J/\lambda^<$ times a function
decaying superpolynomially in $\lambda^< t_{max}$.

However, the lemma then follows by the assumption of $(L/8,\gub)$ unique bulk state.
\end{proof}
\end{lemma}

We need an estimate of the Berry phase.  This is similar to the ideas in\cite{berry1}.
Above, we claimed that
\be
U|\Psi_0\rangle=z |\Psi_0 \rangle,
\ee
for some $z$ with $|z|=1$.
We now determine the value of $z$.  We claim that
\begin{lemma}
\be
z=\exp(-i 2\pi \overline Q_X),
\ee
where
\be
\overline Q_X=\langle \Psi_0 | Q_X | \Psi_0\rangle.
\ee
\begin{proof}
By Eq.~(\ref{Berryprop}), 
\be
\label{fiXc}
\langle \Psi_s| {\cal D}_s | \Psi_s \rangle=0.
\ee
However, since ${\cal D}_s$ an exact quasi-adiabatic evolution operator
\be
i{\cal D}_s |\Psi_s\rangle=\partial_s\Psi_s=i(Q_X-c)\Psi_s,
\ee
for some constant $c$.  Eq.~(\ref{fiXc}) lets us determine $c$ so that
$i{\cal D}_s |\Psi_s\rangle=(Q_X-\overline Q_X)\Psi_s$.
Thus,
$U|\Psi_0\rangle=\exp(i 2\pi Q_X) \exp(-i2\pi \overline Q_X) \Psi_0=
\exp(-i2\pi \overline Q_X) \Psi_0$.
\end{proof}
\end{lemma}

We now consider the phase $z_1$ in lemma (\ref{l8}).
Consider a flux insertion operator $W_1(x_0)$ defined precisely as the above $W_1$ was defined above, except
for inserting the flux along the line $x=x_0$, rather than along the line $x=0$.
That is, we define
\be
Q_X(x_0) =\sum_{i}^{1+x_0\leq x(i)\leq L/2} q_i,
\ee
and
\be
H(\theta_1,\theta_2,x_0)=\sum_Z H_Z(\theta_1,\theta_2,x_0),
\ee
where $H_Z(\theta_1,\theta_2,x_0)$ is defined as follows.  
If the set $Z$ is within distance $R$ of
the vertical line $x=x_0$, then $H_Z(\theta_1,\theta_2)=\exp(i \theta_1 Q_X) H_Z \exp(-i \theta_2 Q_X)$;
if the set $Z$ is within distance $R$ of
the vertical line $x=L/2$, then $H_Z(\theta_1,\theta_2)=\exp(i \theta_2 Q_X) H_Z \exp(-i \theta_2 Q_X)$;
otherwise, $H_Z(\theta_1,\theta_2)=H_Z$.
Then, we can define ${\cal D}_s^{(1)}(x_0)$ and $W_1(x_0)$ similarly to before.
Note that $W_1(0)=W_1$.

For any $x$, define
\be
\overline Q_{x}=\sum_{i,x(i)=x} \langle \Psi_0 | q_i | \Psi_0 \rangle.
\ee
Then, we claim that
\begin{lemma}
The quantity
\be
|\langle \Psi_0 | W_1(0)^{\dagger} W_1(1) | \Psi_0 \rangle-\exp(-i 2\pi \overline Q_{1})|
\ee
is 
bounded by 
$J/\gamma$ times a function decaying superpolynomially in $L$ plus ${\rm poly}(L) J/\lambda^<$ times a function
decaying superpolynomially in $\lambda^< t_{max}$.
\begin{proof}
We have
\be
\langle \Psi_0 | W_1(0)^{\dagger} W_1(1) | \Psi_0 \rangle=
\langle \Psi_0 | W_1(0)^{\dagger} W_2^\dagger W_2 W_1(1) | \Psi_0 \rangle.
\ee

By the above lemma,
$W_2 W_1(0) | \Psi_0 \rangle$ is superpolynomially close to $\Psi_0$ times some phase $z(0)$.  Similarly, one can prove
$W_2 W_1(1) | \Psi_0 \rangle$ is superpolynomially close to $\Psi_0$ times some phase $z(1)$; note that the support of the operators
$W_1(1)$ and $W_2$ still do not overlap.
Thus, 
\be
\langle \Psi_0 | W_1(0)^{\dagger} W_1(1) | \Psi_0 \rangle
\ee
is superpolynomially close to $z(0)^\dagger z(1)$.

However, by the above lemma,
$z(0)$ is superpolynomially close to $\exp(-i 2\pi \sum_{x=1}^{L/2} \overline Q_x)$ and
$z(1)$ is superpolynomially close to $\exp(-i 2\pi \sum_{x=2}^{L/2} \overline Q_x)$.
So,
$z_1(0)^\dagger z_1(1)$ is superpolynomially close to $\exp(-i 2\pi \overline Q_1)$.
\end{proof}
\end{lemma}

Note that the choice of the line $x=0$ was arbitrary.  We can pick any line $x=x_0$ to insert flux into, and in that
way define an operator $W_1(x_0)$, and then we can choose another line  $x=x_0+L/2$ to define $W_2(x_0)$ and use those two lines and repeat
the above proofs, showing that
the quantity
\be
|\langle \Psi_0 | W_1(x_0)^{\dagger} W_1(x_0+1) | \Psi_0 \rangle-\exp(-i 2\pi \overline Q_{x_0+1})|
\ee
is 
bounded by 
$J/\gamma$ times a function decaying superpolynomially in $L$ plus ${\rm poly}(L) J/\lambda^<$ times a function
Thus, for any $x_0$ we have
\begin{corollary}
Assume that the system has an $(L/8,\gub)$ unique bulk state, with $\gub$ greater than or equal to $1/{\rm poly}(L)$.
Assume that the system has a mobility gap with $t_{max}$ superpolynomially large in $L$.  Then,
\be
{\rm min}_{z_1,|z_1|=1}\Bigl( ||W_1(x_0) |\Psi_0 \rangle- z_1 |\Psi_0\rangle| \Bigr)
\ee
is bounded by 
$J/\gamma$ times a function decaying superpolynomially in $L$ plus ${\rm poly}(L) J/\lambda^<$ times a function
decaying superpolynomially in $\lambda^< t_{max}$ and also
\be
\langle \Psi_0 | W_1(x_0) | \Psi_0 \rangle
\ee
is superpolynomially close to 
\be
z \exp(-i 2\pi \sum_{x=0}^{x_0} \overline Q_{x}),
\ee
for some $z$ which is independent of $x_0$.
\end{corollary}

Thus, unless $\overline Q_x$ is an integer for all $x$,
we have defined a flux operator which has an expectation value which depends on the particular line $x_0$ we choose.  If we considered the case of a translation
invariant system, this would be a contradiction, and would prove
that there is not a mobility gap and a unique ground state (this is how
the proof of the higher-dimensional Lieb-Schultz-Mattis theorem goes).  In
this case, we simply identify that there is a position-dependent
expectation value of a flux insertion operator.

\section{Lieb-Robinson Bounds for Quasi-Adiabatic Continuation}
The previous section relied on the fact that quasi-adiabatic evolution can be approximated by a sum of local operators, and
hence the evolution ${\cal D}$ could be approximated to superpolynomnial accuracy by a sum of ${\cal D}^{(1)}+{\cal D}^{(2)}$.
We now want to consider a stronger property.  Suppose we have an operator, such as ${\cal D}$ which can be approximated by
a sum of local operators.  For example, suppose ${\cal D}$ can be approximated to superpolynomial accuracy, in $l$, by a sum of operators
${\cal D}^Z$
supported on sets $Z$ of diameter at most $l$.  Then, {\it if we had an the additional bound} on $\Vert {\cal D}^Z \Vert$,
then we would have a Lieb-Robinson bound for the unitary evolution
\be
U_s={\cal S}' \exp\{i\int_0^s {\rm d}s' {\cal D}_{s'}\}.
\ee

In the next two sections, on Hall conductance, we will rely on an assumption of a Lieb-Robinson bound for the quasi-adiabatic evolution
operator.  This bound is fairly immediate to prove if we consider the slightly simpler case of a mobility gap rather than a spectral
gap.  For example, consider the operator defined in Eq.~(\ref{uncorr}).  For simplicity, we can use an exact quasi-adiabatic
evolution operator.  Then, we have a bound on the operator ${\cal D}(H_s,\partial_s H_Z(s))$ bounded in norm by a constant
times $(1/\gamma) \Vert \partial_s H_Z(s) \Vert$, and the operator also decays superpolyomially in space.  Hence, we have a Lieb-Robinson bound(see the appendix for more discussion of this case).

However, in the case of a corrected quasi-adiabatic continuation operator, we also have to worry about the contribution from
states below the mobility gap.  In this case, the contribution of these states, the term
$\int F(\lambda^< t) W_{\gamma,G}^{loc}(O,t) {\rm d}t$ in the corrected quasi-adiabatic evolution operator, is local as in lemma (\ref{locallemma}),
but the bound on the norm is quite weak.  In particular, the norm may scale with $1/\lambda^<$, and hence may scale
polynomially with $L$.  This makes it difficult (perhaps impossible) to directly prove the Lieb-Robinson bound for corrected
quasi-adiabatic evolution directly from the assumption of a mobility gap, so we will need one additional assumption.

In this section, we  will define the property of a Lieb-Robinson bound for corrected quasi-adiabatic evolution that we need in the next
two sections.  We also present one simple assumption on the local density of states under which this Lieb-Robinson bound can be
derived.  Then, the results  in the next two sections will depend either on the assumption of the Lieb-Robinson bound for
corrected quasi-adiabatic evolution, or on the (fairly mild) assumption on the density of states.

We will consider parameter-dependent Hamiltonians in the next two sections where flux is inserted along lines.  We will consider, however,
more lines than in the previous section.  In the case of a torus, we will have two lines describing flux inserted in one direction (the
``horizontal" direction of the torus) and two other lines describing flux inserted in the other direction (the ``vertical" direction of
the torus).  The reason we have two lines in each direction is similar to the case in the above section: we use the fact that if
we insert opposite flux on two different lines then the Hamiltonian is only changed by a unitary transformation.  The reason we need
two different directions is that the Hall conductance is equal to the curvature when transported around an infinitesimal loop in flux space.

We will need the following Lieb-Robinson bound:
\begin{definition}
Consider a particular parameter dependent Hamiltonian $H_s$.  Consider the quasi-adiabatic evolution operator, ${\cal D}_s$ at $s=0$, and
corresponding unitary $U_s$.
Then, we say that two sets $A,B$ are {\bf separated} if, for $s$ of order unity, we have that for any operator $O_A$ supported on $A$
and $O_B$ supported on $B$ that
$\Vert [U_s O_A U_s^\dagger,O_B]\Vert$ and
$\Vert [U_s^\dagger O_A U_s,O_B]\Vert$ are both bounded by
$|A| \Vert O_A \Vert \Vert O_B \Vert$ times a quantity superpolynomially small in ${\rm dist}(A,B)$.
\end{definition}

This Lieb-Robinson bound can be proven, as explained above, for a system with a spectral gap.  If there is a mobility gap, it can be
proven under the following assumption (with slight modification in the definition of the corrected quasi-adiabatic continuation operator).  We will consider later parameter-dependent Hamiltonians such that $\partial_s H_s=\sum_Z \partial_s H_Z(s)$ is supported on the sets of sites within distance $R$ of two vertical lines (the solid and dashed vertical lines in Figs.~2 and 3).
The derivatives $\partial_s H_Z(s)$ will be non-vanishing only for sets $Z$ which are within distance $R$ of one of these lines.
We will be interested in sets $A$ and $B$ which are connected by part of one of these lines.

Consider a given set $Z$.  The operator $\partial_s H_Z(s)$ for that set has matrix elements between the ground state and various
excited states.  We can define a density of states, $\rho_Z(E)$ by
\be
\rho_Z(E)=\Bigl| (1-P_E) \Bigl( \partial_s H_Z(s) \Bigr) | \Psi_0 \rangle \Bigr|^2,
\ee
where $P_E$ projects onto states with energy $E_0+E$ or more and the partial derivatives are taken at $s=0$ (later when we use
this density of states, we will always be considering Hamiltonians which are unitarily equivalent for different $s$, so the density of states $\rho_Z(E)$ will be independent of $s$).  

While it is expected that a disordered system will have states with energy of order $1/V$, we expect that only for $Z$ close
to certain points will $\partial_s H_Z(s)$ have non-negligible matrix elements to these states.  In contrast, most $Z$ are expected to have the property
that $\partial_s H_Z(s)$ will only produce non-negligible matrix
elements from the ground state to excited states
with energies of order unity.  In fact, we will require even weaker conditions than that.  We will allow a typical $Z$ to have non-negligible
matrix elements to energy which are of order $1/L^{\alpha}$, for $\alpha<1$.

Suppose there is an energy $\Delta \sim \gamma/L^\alpha$ such that the following property holds.  We define $\Slow$ to be the set of $Z$ for which
$\rho(\Delta)$ is bounded by $\rho_{max}$ times $\Vert \partial_s H_Z(s) \Vert$, where
$\rho_{max}$ is a quantity which is superpolynomially small in $L$.  We define
$\Fast$ to be the remaining $Z$.  Then, we define the corrected quasi-adiabatic continuation operator
as
\be
{\cal D}_s=\sum_Z {\cal D}_s^Z
\ee
as in Eq.~(\ref{de}).
However,
for $Z\in \Slow$, we define ${\cal D}_s^Z$ with the cutoff $\lambda^<=\Delta$, while for $Z \in \Fast$ we use the cutoff
$\lambda^<=\lambda_{min}$.

Then, this operator ${\cal D}_s$ continues to approximate the exact evolution up to superpolynomially small error bounded
by $(\rho_{max}/\lambda^<) \sum_{Z\in \Slow} \Vert \partial_s H_Z(s) \Vert$, with
the additional
error due to corrections from states below energy $\Delta$.

Now, this operator ${\cal D}_s$ will separate sets $A$ and $B$ under mild assumptions on the density of states $\rho_Z(E)$.
The operators ${\cal D}_s^Z$ for $Z\in \Fast$ will have large norm, but will decay exponentially in space.  Assume $\lambda_{min}\geq {\rm poly}(1/L)$.  If we can find
a segment of the line of length separating sets $A,B$ as shown in Fig.~1.
which scales as $L^\beta/\gamma$, for some $\beta$, with $\alpha<\beta \leq 1$, such that all $Z$ in that segment
are in $\Slow$, then we will have the desired Lieb-Robinson bound:
the $Z\in\Slow$ will give operators
${\cal D}_s^Z$ bounded in norm by
${\cal O}((L^\alpha/\gamma) \Vert \partial_s H_Z(s) \Vert)$, and so have
a Lieb-Robinson velocity.
 the $Z\in \Fast$ will have some effect on the dynamics in this segment,
due to long-distance tails of $\partial_s H_Z(s)$ (i.e., even is $Z\in \Fast$,
the operator $\partial_s H_Z(s)$ has support in $\Slow$); however, this
produces only corrections of order
${\rm poly}(L)$ times a quantity decaying exponentially in $L^\alpha$.  
Hence, we will have the desired separation.

\begin{figure}
\centering
\includegraphics[width=200px]{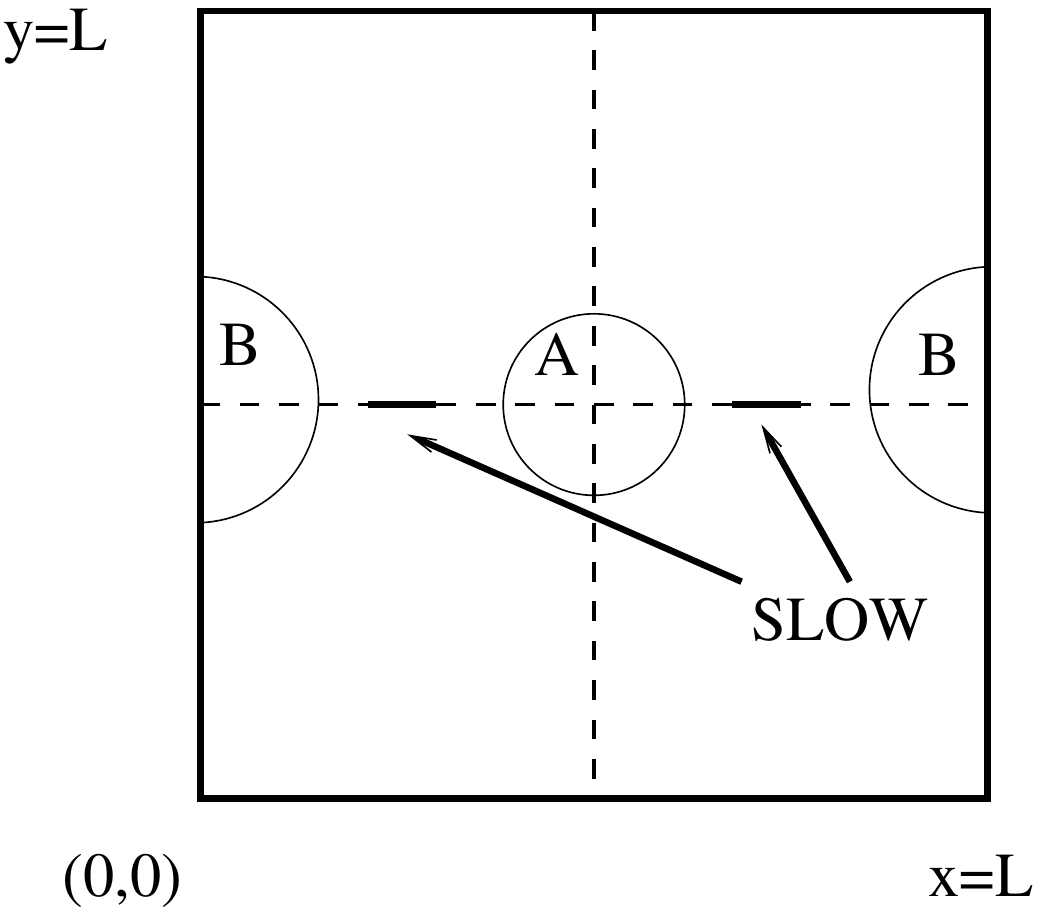}
\caption{{\small{Line illustrating an example of separation on the torus.  Sets $A$ and $B$ are marked by circles; there are periodic
boundary conditions so $B$ is a contiguous set.  The flux is inserted along the dashed line connecting them.  The short solid intervals
mark sets $\Slow$ separating $A$ and $B$ as described.
}}}
\end{figure}

\section{Hall Conductance on a Torus}
We now prove Hall conductance on a torus.  For a site $i$, we define it to have $x$ and $y$ coordinates $x(i)$ and $y(i)$.
We consider a Hamiltonian with a mobility gap, with $t_{max}$ superpolynomially large in $L$.  We now define
We define a parameter-dependent
Hamiltonian
$H(\theta_1,\theta_2,\phi_1,\phi_2)$ as follows.  We pick two vertical lines at $x_1,x_2$ and two horizontal lines $y_1,y_2$ to insert
flux.
Let $Q_X$ be defined by
\be
Q_X =\sum_{i}^{x_1\leq x(i)\leq x_2} q_i,
\ee
where $x(i)$ is the $\hat x$-coordinate of site $i$ and
\be
Q_Y =\sum_{i}^{y_1\leq y(i)\leq y_2} q_i.
\ee
We fix $|x_1-x_2|=|y_1-y_2|=L/2$.
\begin{definition}
\label{twTdef}
Let $H$ be any operator which can be written as $H=\sum_Z H_Z$ with the $H_Z$ supported on a set $Z$ of diameter less than
$L/2$.  Assume that all the sets $Z$ are squares.
Then, each such $H_Z$ intersects at most one of the lines $x=x_1$ or $x=x_2$ and at most one of the lines $y=y_1$ or $y=y_1$.
Then, define the {\bf twisted operator}
\be
H(\theta_1,\theta_2,\phi_1,\phi_2)=\sum_Z H_Z(\theta_1,\theta_2,\phi_1,\phi_2),
\ee
where $H_Z(\theta_1,\theta_2,\phi_1,\phi_2)$ is defined as follows.
If the set $Z$ intersects
the vertical line $x=x_1$, then $H_Z(\theta_1,\theta_2,\phi_1,\phi_2)=\exp(i \theta_1 Q_X) H_Z(0,0,\phi_1,\phi_2) \exp(-i \theta_1 Q_X)$;
if the set $Z$ intersects
the vertical line $x=x_2$, then $H_Z(\theta_1,\theta_2,\phi_1,\phi_2)=\exp(-i \theta_2 Q_X) H_Z(0,0,\phi_1,\phi_2) \exp(i \theta_2 Q_X)$;
otherwise $H_Z(\theta_1,\theta_2,\phi_1,\phi_2)=H_Z(0,0,\phi_1,\phi_2)$.
If the set $Z$ intersects
the horizontal line $y=y_1$, then $H_Z(0,0,\phi_1,\phi_2)=\exp(i \phi_1 Q_Y) H_Z \exp(-i \phi_1 Q_Y)$;
if the set $Z$ intersects
the horizontal line $y=y_2$, then $H_Z(0,0,\phi_1,\phi_2)=\exp(-i \phi_2 Q_Y) H_Z \exp(i \phi_2 Q_Y)$;
otherwise $H_Z(0,0,\phi_1,\phi_2)=H_Z$.
\end{definition}
Note that we chose the sets $Z$ to be squares so that they would be contiguous sets; thus, if $Z$ was close to a line $x=x_1$ and contained
some points with $x<x_1$ and some with $x>x_1$ then $Z$ will intersect the line $x=x_1$, and similarly for the other three lines.
This definition defines our parameter-dependent Hamiltonian, but we will also use it later for other operators.

\begin{figure}
\centering
\includegraphics[width=200px]{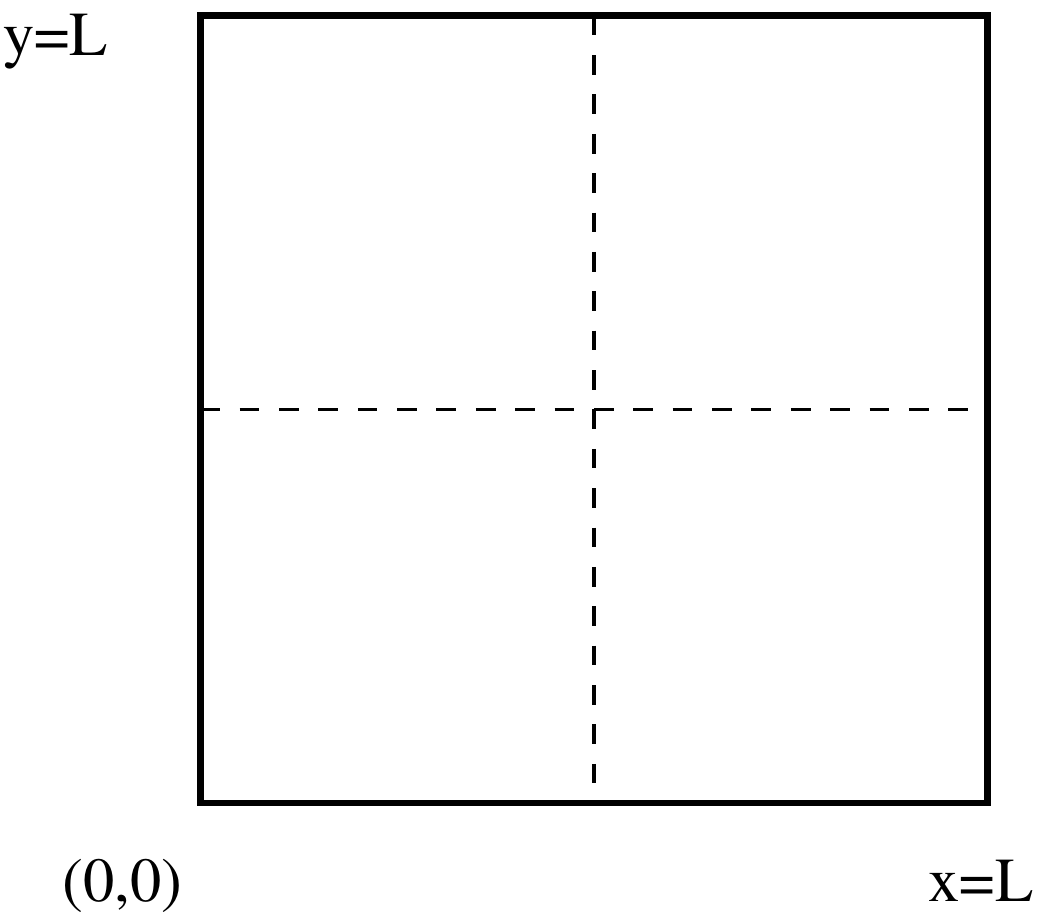}
\caption{{\small{Lines illustrating how the twists are defined on the torus.  The twists $\theta_1,\phi_1$ affect interactions
close to the vertical and horizontal solid lines, respectively, while the twists $\theta_2,\phi_2$ affect interactions
close to the vertical and horizontal dashed lines.
}}}
\end{figure}

In this section, we prove the following:
\begin{theorem}
Consider a Hamiltonian with a mobility gap, with $t_{max}$ superpolynomially large in $L$,
and $\lambda_{min}$ greater than or equal to $1/{\rm poly}(L)$.
Suppose we can find lines $x_1,x_2,y_1,y_2$ with $|x_1-x_2|=L/2$ and $|y_1-y_2|=L/2$
such that the following holds.  Consider any pair of lines.  Let $A$ be the set of points within
distance $L/8$ of the intersection of that pair of lines.  Consider any other pair of lines.
Let $B$ be the set of points within distance $L/8$ of the intersection of that pair of lines.  Consider quasi-adiabatic evolution
under any of the four different quasi-adiabatic evolution
operators ${\cal D}^{\theta,1}, {\cal D}^{\theta,2}, {\cal D}^{\phi,1}, {\cal D}^{\phi,2}$
defined below.
Suppose that $A$ and $B$ are separated under quasi-adiabatic evolution under 
all four such operators,
Then, the Hall conductance is quantized to $n(e^2/h)$,  for some integer $n$, up to an error which is superpolynomially small in $L$.
\end{theorem}

Note that by assumption the Hamiltonian has an
a $(R,\gub)$ unique ground state, with $\gub$ greater than or equal to $1/{\rm poly}(L)$.
We only sketch the proof, since it essentially follows \cite{hall} (this proof has some similarity with the Chern
number approach\cite{av} but avoids any averaging assumptions).  We pick a parameter $r$ which is superpolynomially small in $L$ (the
choice of $r$ will be given later),
with $2\pi/r=N$ for some integer $N$.
Following \cite{hall},
we define a set of set of $N^2$ different closed paths in parameter space.  Each path keeps $\theta_2=\phi_2=0$ throughout.  Each
path is defined by a given pair of numbers $(\theta_x,\theta_y)$ with $\theta_x=m r$ and $\theta_y=n r$ for a pair of integers
$m,n$ in the range $0,...,N-1$. 
The paths start at $\theta_1=\phi_1=0$.  Then, we move to $\theta_1=0,\phi_1=\theta_y$.  Then we move to $\theta_1=\theta_x,\phi_1=\theta_y$.
Then we move around a small square loop of size $r$ at $\theta_x,\theta_y$ as follows: we move to
$\theta_1=\theta_x+r,\phi_1=\theta_y$, to
$\theta_1=\theta_x+r,\phi_1=\theta_y+r$,
$\theta_1=\theta_x,\phi_1=\theta_y+r$,
and back to
$\theta_1=\theta_x,\phi_1=\theta_y$.  Finally, we return to the origin by moving to
$\theta_1=0,\phi_1=\theta_y$ and then to $\theta_0=\phi_1=0$.

Thus, each path consists of $8$ different line segments.  We break it into three distinct parts.  First,
$V(m,n)$ moving from $0,0$ to 
to $\theta_1=0,\phi_1=\theta_y$ and then to $\theta_1=\theta_x,\phi_1=\theta_y$.  Then,
$V_\circlearrowleft(m,n)$ moving around the small square loop.  Finally, $V^\dagger(m,n)$  returning to the origin.
We decompose the first part, $V(m,n)=U_2(m,n) U_1(m,n)$, with $U_{1}(m,n),U_2(m,n)$ corresponding to the two segments of
that part.  We also define unitaries $\tilde U_1(m,n),\tilde U_2(m,n)$ corresponding to evolution along the
same path of  $\theta_1,\phi_1$ but with $\theta_2=-\theta_1$ and $\phi_2=-\phi_1$.
We also decompose the motion around the square loop as
$V_{\circlearrowleft}(m,n)=U_{D}(m,n) U_L(m,n) U_U(m,n) U_R(m,n)$ corresponding to moving around the four different
sides of the square (we use subscripts $D,L,U,R$ indicating the the motion is first to the right, then up, then to the left,
then down).  We also define $\tilde U_D(m,n),\tilde U_L(m,n),\tilde U_U(m,n),\tilde U_R(m,n)$ corresponding to
evolution around the same square but with $\theta_2=-\theta_1$ and $\phi_2=-\phi_1$.

Given these paths, we have to pick a quasi-adiabatic evolution operator to generate the various unitaries $U$ and $\tilde U$.
We define ${\cal D}^{\theta}(\theta,\phi)$ to be the quasi-adiabatic
continuation operator which produces an infinitesimal
change in $\theta$ starting at $\theta_1=-\theta_2=\theta$ and $\phi_1=-\phi_2=\phi$.  That is, for $H_s=H(\theta+s,-\theta-s,\phi,-\phi)$, we
define ${\cal D}^{\theta}(\theta,\phi)={\cal D}_s$ at s=0.
We define ${\cal D}^{\phi}(\theta,\phi)$
to
produce an infinitesimal
change in $\phi$, again starting at $\theta_1=-\theta_2=\theta$ and $\phi_1=-\phi_2=\phi$.  
We use these operators ${\cal D}^{\theta}(\theta,\phi),{\cal D}^{\phi}(\theta,\phi)$ to generate the unitaries
$\tilde U_1(m,n),\tilde U_2(m,n),\tilde U_{D,L,U,R}(m,n)$.

Let us consider the operator ${\cal D}^{\theta}(0,0)$.  For
$\theta_1=-\theta_2=s$ and $\phi_1=-\phi_2=0$,
we have that $\partial_s H(s)=\sum_Z \partial_s H_Z(s)$, and
$\partial_s H_Z(s)$ is nonvanishing if $Z$ is within distance $R$ of the line at $x=x_1$ or if $Z$ is within
distance $R$ of the line at $x=x_2$.
Let $O^{(1)}(s)$ denote the sum of terms in $\partial_s H(s)$ near the line at $x=x_1$ and let
$O^{(2)}(s)$ denote the sum of terms in $\partial_s H(s)$ near the line at $x=x_2$, so that
${\cal D}_s={\cal D}(H_s,O^{(1)}(s))+
{\cal D}(H_s,O^{(2)}(s))$.
Under the assumption of a mobility gap, because of the superpolynomial localizability property,
up to superpolynomially small error. 
we can approximate the operators ${\cal D}(H_s,O^{(1)})$ and ${\cal D}(H_s,O^{(2)})$ by operators ${\cal D}^{\theta,(1)}$ and
${\cal D}^{\theta,(2)}$ supported within distance less than $L/8$ of the respective lines
$x=0$ and $x=L/2$ and which are a sum of squares of operators on squares of size at most $L/2$.
Then, we can apply the definition (\ref{twTdef}) to define 
\be
{\cal D}^{\theta,1}(\theta,\phi)={\cal D}^{\theta,(1)}(\theta,0,\phi,0),
\ee
and
\be
{\cal D}^{\theta,2}(\theta,\phi)={\cal D}^{\theta,(2)}(0,-\theta,\phi,0).
\ee
Note that
\be
\label{dcmpt}
{\cal D}^{\theta}(\theta,0)-
\Bigl( {\cal D}^{\theta,1}(\theta,0)
+{\cal D}^{\theta,2}(\theta,0) \Bigr)
\ee
is superpolynomially small.
We define ${\cal D}^{\phi,(1)}$ and ${\cal D}^{\phi,(2)}$ in an analogous way by decomposing the operator ${\cal D}^{\phi}(0,0)$, and then
we have that
\be
\label{dcmpp}
{\cal D}^{\phi}(0,\phi)-
\Bigl( {\cal D}^{\phi,1}(0,\phi)
+{\cal D}^{\phi,2}(0,\phi) \Bigr)
\ee
is superpolynomially small.
We will use these operators ${\cal D}^{\theta,1}(\theta,\phi)$ and ${\cal D}^{\phi,1}(\theta,\phi)$ to generate the unitaries
$U_1(m,n),U_2(m,n),U_{D,L,U,R}(m,n)$.

The reader will notice one fact about this choice of quasi-adiabatic evolution
operators: they are all defined in terms of
operators at $\theta=\phi=0$ by applying a twist.
This implies
that the separation assumption can be defined solely in terms of the
operators ${\cal D}$ at $\theta,\phi=0$.  For example,
\be
{\cal S}'\exp\{i\int_0^s {\rm d}s' {\cal D}^{\theta,1}(s,0)\}=
\exp(i Q_X s')
{\cal S}'\exp\{i\int_0^s {\rm d}s' \Bigl( -Q_X+{\cal D}^{\theta,1}(0,0)\Bigr)\}
\exp(-i Q_X s').
\ee

The proof rests on the following four facts.  We sketch each in turn.
First, for $r$ sufficiently small, the evolution around the path $U(0,0)$ returns the ground state to the ground state up to
superpolynomial small error and up to some phase which is
that is, up to superpolynomially
small error, equal to $r^2/2\pi$ times the Hall conductance in units of $e^2/h$, plus corrections of
order $r^3$.  
That is, we claim that
\begin{lemma}
\label{spcl}
The quantity 
\be
\label{spclo}
|\langle \Psi_0, U(0,0) \Psi_0 \rangle |
\ee
is superpolynomially close to unity for any $r$ of order unity.  Also, the quantity
\be
\label{or3}
|\langle \Psi_0, U(0,0) \Psi_0 \rangle - \exp(ir^2 \frac{\sigma_{xy}}{2\pi}\frac{h}{e^2})|
\ee
is bounded by terms of order ${\rm poly}(L) {\cal O}(r^3)$ plus terms
which are superpolynomially close to unity.
\begin{proof}
Eq.~(\ref{or3}) follows from a power series expansion of the quasi-adiabatic continuation operator.  The terms of order $r^2$
in the expectation value in Eq.~(\ref{or3}) arise from approximating, along each leg of the square, the
evolution
\be
U_s={\cal S}' \exp\{i\int_0^s {\cal D}_s {\rm d}s'\} = 1+is{\cal D}_0 + {\cal O}(s^2).
\ee
The quasi-adiabatic evolution operator $i{\cal D}_0$ acting on the ground state is superpolynomially close to the derivative
of the ground state with respect to the parameter $s$.  Thus, to order $r^2$, the
expectation value
$\langle \Psi_0, U(0,0) \Psi_0 \rangle$ is superpolynomially close to
$1+r^2 2 {\rm Im} (\langle \partial_\phi \Psi(\theta,\phi), \partial_\theta \Psi(\theta,\phi) \rangle)$,
where $\Psi(\theta,\phi)$ is the ground state of $H(\theta,0,\phi,0)$, and partial derivatives are taken
at $\theta=\phi=0$.
However, this quantity
$2 {\rm Im} (\langle \partial_\phi \Psi(\theta,\phi), \partial_\theta \Psi(\theta,\phi) \rangle)$ is related to
the Hall conductance by the Kubo formula, so Eq.~(\ref{or3}) follows to order $r^2$.  Using a bound on the norm
of ${\cal D}$, we bound the terms of order $r^3$ by ${\rm poly}(L)$.

To show Eq.~(\ref{spclo}), we show that the state $U(0,0)\Psi_0$ has an energy which is superpolynomially close
to $E_0$.  Then,  using the assumed $\lambda_{min}$, Eq.~(\ref{spclo}) will follow.  
We will estimate the energy by estimating the expectation value of $H_Z$ for each $Z$ in the state $U(0,0)\Psi_0$.

Suppose first that $Z$ is far from the point $x=x_1,y=y_1$ where two lines intersect 
(we use the word ``far" to indicate something is a distance of order $CL$ from something, and near otherwise).
Note that $U(0,0)=V_\circlearrowleft(0,0)$.  
We have
$V_\circlearrowleft(0,0)=U_D(0,0) U_L(0,0) U_U(0,0) U_R(0,0)$.  
For simplicity, in this lemma we write $U_L=U_L(0,0),U_R=U_R(0,0)$, and so on.
We claim that the operator $V_{\circlearrowleft}(0,0)$ can be approximated to
superpolynomial accuracy by an operator supported near $x=x_1,y=y_1$ using the separation assumption and superpolynomial localizability property.
So, if $Z$ is far, from $x=x_1,y=y_1$, then
$\Vert [H_Z,U(0,0)]\Vert$ is superpolynomially small, so
$\langle \Psi_0, U(0,0)^\dagger H_Z U(0,0) \Psi_0 \rangle$ is superpolynomially close to
$\langle \Psi_0, H_Z \Psi_0 \rangle$.

We now show that
$V_{\circlearrowleft}(0,0)$ can indeed be approximated by an operator supported near $x=x_1,y=y_1$.  
We will make repeated use of the basic identity for any two Hermitian operators $a_s,b_s$ that,
defining
\be
u_{s_1,s_2}={\cal S}' \exp\{i\int_{s_1}^{s_2} {\rm d}s' b_s\}
\ee
we have
\begin{eqnarray}
\label{basicid}
{\cal S}' \exp\{i\int_0^s {\rm d}s' a_{s'} \} &= &
{\cal S}' \exp\{i\int_0^s {\rm d}s' u_{s',s} (a_{s'}-b_{s'}) u_{s',s}^\dagger \} u_{0,s} \\ \nonumber
&=&
u_{0,s}
{\cal S}' \exp\{i\int_0^s {\rm d}s' u_{s',0}^\dagger (a_{s'}-b_{s'}) u_{s',0} \}.
\end{eqnarray}
Decomposing $V_\circlearrowleft(0,0)$, we have 
\be
U_D U_L U_U U_R =U_D U_U {\cal S}' \exp\{-i U_U^\dagger \int_0^r {\rm d}s' {\cal D}^{\theta,1}(r-s',r) U_U \} U_R
\ee
So, it suffices to show that $U_D U_U$ can be approximated by an operator supported near $x=x_1,y=y_1$ and similarly for
${\cal S}' \exp\{-i U_U^\dagger \int_0^r {\rm d}s' {\cal D}^{\theta,1}(r-s,r) U_U \} U_R$.
We have 
\be
U_D U_U={\cal S}' \exp\{ -i\int_0^r {\rm d}s' {\cal D}^{\phi,1}(0,r-s')\}
{\cal S}' \exp\{ i\int_0^r {\rm d}s' {\cal D}^{\phi,1}(r,s')\}.
\ee
Define $A(s')={\cal D}^{\phi,1}(0,s')-{\cal D}^{\phi,1}(r,s')$.  
Define $W_{s_1,s_2}=
{\cal S}' \exp\{ i\int_{s_1}^{s_2} {\rm d}s' {\cal D}^{\phi,1}(r,s')\}$, so that $U_U=W_{0,r}$.
Then,
\be
U_D U_U={\cal S}' \exp\{ -i\int_0^r {\rm d}s' W_{s',r} A(r-s') W_{s',r}^\dagger \}.
\ee
Note that $A(s')$ is can be approximated by an operator supported near $x=x_1,y=y_1$ using the
superpolynomially localizability property, and so using the separation assumption,
$W_{s',r} A(r-s') W_{s',r}^\dagger$ can also be approximated by such  an operator to superpolynomial
accuracy.  Thus, $U_D U_U$ can be approximated by  such an operator.
We next consider
${\cal S}' \exp\{-i U_U^\dagger \int_0^r {\rm d}s' {\cal D}^{\theta,1}(r-s',r) U_U \} U_R$.
Define 
\be
B(s)
=
U_U^\dagger {\cal D}^{\theta,1}(r-s,r) U_U-
{\cal D}^{\theta,1}(r-s,r),
\ee
and define $Y_{s_1,s_2}={\cal S}'\exp\{-i \int_{s_1}^{s_2} {\rm d}s' {\cal D}^{\theta,1}(s-s',r)\}$ so that $U_L=Y_{0,r}$.
Then,
\be
{\cal S}' \exp\{-i U_U^\dagger \int_0^r {\rm d}s' {\cal D}^{\theta,1}(r-s,r) U_U \} U_R
=
{\cal S}' \exp\{-i \int_0^r {\rm d}s' Y_{s',r} B(s') Y_{s',r}^\dagger \}
U_L U_R.
\ee
Note that $B(s')$ is can be approximated by an operator supported near $x=x_1,y=y_1$ using the
superpolynomially localizability property, and so using the separation assumption,
$Y_{s',r} B(s') Y_{s',r}^\dagger$ can also be approximated by such  an operator to superpolynomial
accuracy.  Thus, 
${\cal S}' \exp\{-i \int_0^r {\rm d}s' Y_{s',r} B(s') Y_{s',r}^\dagger \}$
can also be approximated by such an operator.
Finally, one may show  that $U_L U_R$ can be approximated by such an operator in a similar way as to
the proof that $U_D U_U$ could be approximated by such an operator.

Now, suppose that $Z$ is close to both $y=y_1$ and $x=x_1$.
Consider the expectation value
$\langle \Psi_0, U_R^\dagger U_U^\dagger U_L^\dagger U_D^\dagger 
H_Z U_D U_L U_U U_R \Psi_0 \rangle$.
Note that
$\tilde U_D^\dagger H_Z \tilde U_D$ is superpolynomially close to
$U_D^\dagger H_Z U_D$.  So, we may consider
$\langle \Psi_0, U_R^\dagger U_U^\dagger U_L^\dagger \tilde U_D^\dagger 
H_Z \tilde U_D U_L U_U U_R \Psi_0 \rangle$.
Using the separation assumption, $\tilde U_D^\dagger H_Z \tilde U_D$ has small commutator with
${\cal D}^{\theta,2}$.  So,
$U_L^\dagger \tilde U_D^\dagger 
H_Z \tilde U_D U_L$ is superpolynomially close to
$\tilde U_L^\dagger \tilde U_D^\dagger 
H_Z \tilde U_D \tilde U_L$ (this calculation uses the identity (\ref{basicid}) and similar decomposition of unitaries into
ordered exponentials as the above case, so we omit the details).
Repeating this, we find that 
$\langle \Psi_0, U_R^\dagger U_U^\dagger U_L^\dagger U_D^\dagger 
H_Z U_D U_L U_U U_R \Psi_0 \rangle$ is superpolynomially close to
$\langle \Psi_0, \tilde U_R^\dagger \tilde U_U^\dagger \tilde U_L^\dagger \tilde U_D^\dagger 
H_Z 
\tilde U_D \tilde U_L \tilde U_U \tilde U_R \Psi_0 \rangle$.
However, $\tilde U_D \tilde U_L \tilde U_U \tilde U_R \Psi_0 \rangle$ is superpolynomially close to
$\Psi_0$ times a phase.  Therefore,
$\langle \Psi_0, U_R^\dagger U_U^\dagger U_L^\dagger U_D^\dagger 
H_Z U_D U_L U_U U_R \Psi_0 \rangle$ is superpolynomially close to
$\langle \Psi_0, 
H_Z \Psi_0 \rangle$.  This completes the possible cases.
\end{proof}
\end{lemma}
Let us discuss one point in the above proof.  This approach used to show Eq.~(\ref{spclo}) is different than the approach used
in \cite{hall}, where it was simply noted that quasi-adiabatic evolution around a closed loop in a gapped region of parameter
space approximates adiabatic evolution and hence returns the ground state to the ground state up to small error; we use this different
approach because here we do not want to rely on any assumption of a mobility gap for parameter values other than $\theta=\phi=0$.
In \cite{hall}, since $r$ was taken superpolynomially small, evolution within the region of parameter space with $|\theta_x|,|\theta_y|\leq r$
preserved the spectral gap assumed in that paper.  In this paper, we could try to follow that idea and show that
a mobility gap at $\theta=\phi=0$ implies a mobility gap for superpolynomially small change in the Hamiltonian; this may be possible,
but since an alternate method is available we preferred not to do this.

The second property is that the evolution around all loops returns the ground state to the ground state and
produces the same expectation value up to superpolynomially small error.
That is,
\begin{lemma}
\label{lemsmae}
Define 
\be
z(m,n)=
\langle \Psi_0,U(m,n) \Psi_0 \rangle.
\ee
Then,
\be
|z_{m,n}-z_{0,0}|
\ee
is superpolynomially small.
\begin{proof}
To prove this, note that the expectation value is equal to
$\langle \Psi_0 | V(m,n)^\dagger V_\circlearrowleft(m,n) V(m,n)| \Psi_0 \rangle$.  The operator
$V_{\circlearrowleft}$ can be approximated to superpolynomial accuracy
by an operator supported near $x=x_1,y=y_1$.
So, 
$U_2(m,n)^\dagger V_{\circlearrowleft}(m,n) U_2(m,n)$ is superpolynomially close to
$\tilde U_2(m,n)^\dagger V_{\circlearrowleft}(m,n) \tilde U_2(m,n)$.  Again using the Lieb-Robinson bounds, 
$U_1(m,n)^{\dagger} U_2(m,n)^\dagger V_{\circlearrowleft}(m,n) U_2(m,n) U_1(m,n) $ is superpolynomially close to
$\tilde U_1(m,n)^\dagger \tilde U_2(m,n)^\dagger V_{\circlearrowleft}(m,n) \tilde U_2(m,n) \tilde U_1(m,n)$.
However, $\tilde U_2(m,n)  \tilde U_1(m,n) | \Psi_0\rangle=\exp(i Q_X \theta_1) \exp(i Q_Y\theta_2) | \Psi_0 \rangle$ and
$V_{\circlearrowleft}(m,n)$ is superpolynomially close to
$\exp(i Q_X \theta_1) \exp(i Q_Y\theta_2) 
V_{\circlearrowleft}(m,n)
\exp(-i Q_X \theta_1) \exp(i Q_Y\theta_2)$.
So, 
$\langle \Psi_0 | U(m,n) | \Psi_0 \rangle $ is superpolynomially close to
$\langle \Psi_0 | U(0,0) | \Psi_0 \rangle $.
\end{proof}
\end{lemma}

The third property is that the product $\Bigl(U(N-1,N-1) U(N-2,N-1) ... U(0,N-1)\Bigr) \Bigl( U(N-1,N-2) .... U(0,N-2) \Bigr)... \Bigl(U(N-1,0 ... U(0,0) \Bigr)$ is exactly equal to the unitary operator $U_{big}$ corresponding to quasi-adiabatic evolution around a big loop of size $2\pi$: moving from
$\theta_1=\phi_1=0$ to $\theta_1=2\pi,\phi_1=0$, to
$\theta_1=\phi_1=2\pi$ to $\theta_1=0,\phi_1=2\pi$, to $\theta_1=\phi_1=0$.  This is the evolution decomposition in \cite{hall}.

The fourth property is that the phase due to  evolution around the big loop is equal to unity up to superpolynomially small error.
To prove this, we need to show that evolution along any of the four sides of the big loop (for example, from $\theta_1=\phi_1=0$ to
$\theta_1=2\pi,\phi_1=0$) send the ground state to the ground state up to a phase, up to superpolynomially small error.
Then, the phase will cancel between opposite sides of the loop.
However, the fact that evolution along a single side sends the ground state to the ground state up to a phase and up to
superpolynomially small error follows from the same argument as in the Lieb-Schultz-Mattis proof previously.
Thus,
\be
\langle \Psi_0,U_{big} \Psi_0 \rangle
\ee
is superpolynomially close to unity.

We then approximate the phase of evolution around the large loop by the product of phases for evolution around each small loop, following
the Appendix in \cite{hall}.  
That is, 
we approximate:
\be
1\approx \langle \Psi_0,U_{big} \Psi_0 \rangle  \approx \prod_{m,n} z_{m,n}.
\ee
The error in the second approximation is due to the fact that
each small loop has some ``leakage" outside the ground state.  That is, $U(m,n) \Psi_0$ has some component perpendicular to the
ground state.  Define $E_{m,n}$ to be the $|(1-P_0) U(m,n) \Psi_0|^2=1-|z_{m,n}|^2$.  Then,
the difference
\be
\label{diffin}
|\langle \Psi_0,U_{big} \Psi_0 \rangle  \approx \prod_{m,n} z_{m,n} | \leq \sum_{m,n} \sqrt{E_{m,n}}.
\ee
However, by lemma (\ref{spcl}), the quantity $E_{0,0}$ is superpolynomially small and so by lemma (\ref{lemsmae}),
the quantity $E_{m,n}$ is superpolynomially small for all $m,n$.
The number of small loops scales as $1/r^2$.  Hence, we can pick an $r$ which is superpolynomially large
such that the difference in Eq.~(\ref{diffin}) is negligible.
Thus,
$|1- \prod_{m,n} z_{m,n}|$ is superpolynoomially small.
Then, since the phase is almost  the same for every small loops, the phase around the loop with $m=n=0$ is close to
$2\pi n/N=2\pi(r/2\pi)^2$ for some integer $n$.  Hence the Hall conductance is close to $ne^2/h$.

Note that we used three properties of these operators ${\cal D}^{\theta}(\theta,\phi)$ and ${\cal D}^{\phi}(\theta,\phi)$.  First, that
each operator can be approximated, to superpolynomial accuracy, by a sum of operator operators
supported close to the respective lines $x=x_1,x_2$ or $y=y_1,y_2$, each of which is approximated by a sum of operators on squares of
increasing size but superpolynomially decaying strength.
Second, the separation assumption.  Third, 
evolution of the ground state under ${\cal D}^{\theta}$ or ${\cal D}^{\phi}$
approximates,
to superpolynomial accuracy, the adiabatic evolution of the ground
state, and hence
gives a state which is 
is superpolynomially close to $\exp(i Q_X \theta) \exp(i Q_Y \phi) \Psi_0$.
Thus, any operators which satisfy these three properties would suffice.

\section{Hall Conductance on an Annulus}
We now consider open boundary conditions.  Specifically, we consider a system on an annulus.  In this case, a system will not have
a gap: there will be gapless edge modes.  However, there is still a bulk gap.  We need to define this.
We let $\In$ and $\Out$ denote the sites on the inner and outer edges of the annulus.  We let $\B$ denote the sites $s$ such that
${\rm dist}(s,I)\geq L/3,{\rm dist}(s,O)\geq L/3$.  That is, $\B$ is a ring around the inside of the annulus, which we
call the ``bulk".  The constant factor
$1/3$ is not particularly important; we simply want a constant less than $1/2$ (so that the widht of the ring scales linearly in
$L$) and greater than $0$ (so that the distance from the edges also scales linearly with $L$).
We use $x$ and $y$ coordinates to label sites, and let the line $y=0$ correspond to halfway between the inner and outer rings
of the annulus, so that $\B$ includes points with $-L/6\leq y \leq L/6$.

\begin{definition}
We say that a system has a {\bf bulk mobility gap} $\gamma$ and localization length $\xi$ and localization constant $\con$ 
up to time $t_{max}$ if,
for any operator $O$ supported on set $A$, there exists an operator $W_{\gamma,G}^{loc}(O,t)$ with 
the following properties.
First, for any $l$, 
$W_{\gamma,G}^{loc}(O,t)$ can be approximated by a sum of operators $P_I+P_O+P_{bulk}$, wth
$P_I$ is supported on $b_l(\In)$, $P_O$ is supported on $b_l(\Out)$, and $P_{bulk}$ is supported on
$b_l(A)$,
up to an error in operator norm bounded by
\be
\label{leak2}
\con\exp(-l/\xi) \Vert W_{\gamma,G}(O) \Vert
+{\rm max}_{|\omega|\geq \gamma} |\tilde G(\omega)|
\Vert O \Vert
\ee
with
\be
\Vert P_I \Vert \leq \Vert O \Vert \con \exp(-{\rm dist}(A,I)/\xi),
\ee
\be
\Vert P_O \Vert \leq \Vert O \Vert \con \exp(-{\rm dist}(A,O)/\xi).
\ee
Second, we
require that the state produced by acting with $W_{\gamma,G}(O)(t)=\exp(i H_0 t) W_{\gamma,G}(O)\exp(-i H_0 t)$ on the ground state
is equal to the state produced by acting with an operator $W_{\gamma,G}^{loc}(O,t)$ on the ground state, i.e.,
\be
W_{\gamma,q}(O)(t) \Psi_0 = W_{\gamma,G}^{loc}(O,t)  \Psi_0.
\ee
Third, we have
\be
\Vert W_{\gamma,G}^{loc}(O,t) \Vert \leq \Vert W_{\gamma,G}(O) \Vert.
\ee
\end{definition}

\begin{definition}
\label{uniquebulk}
We say that a Hamiltonian $H$ has an $(l,\gub)$ {\bf unique bulk ground state} if the following holds.
Given any density matrix $\rho$ such that, for all sets $A\subset b_l(\B)$ with ${\rm diam}(A)\leq l$ we have
\be
\Vert {\rm Tr}_{\overline A}(\rho-P_0) \Vert_1 \leq \epsilon,
\ee
then
\be
\Vert {\rm Tr}_{\overline \B}(\rho-P_0) \Vert_1 \leq \gub \sqrt{\epsilon}
\ee
\end{definition}

The definition of a unique bulk state is certainly necessary.  Let us explain why.  Consider a fractional Hall system defined on an annulus.
The system has a bulk gap.  However, the system does not have a unique bulk state according to our definition: physically, one expects
$q$ different bulk states in a $1/q$ Laughlin wavefunction.  Such a system will also not have an integer Hall conductance.  Thus, to
prove integer quantization we need to make the assumption of a unique bulk state.

We now define twisted boundary conditions.
We pick two vertical lines at $x_1,x_2$ with $|x_1-x_2|=L/2$ and a single
horizontal line at $y=0$.
Let $Q_X$ be defined by
\be
Q_X =\sum_{i}^{x_1\leq x(i)\leq x_2} q_i,
\ee
where $x(i)$ is the $\hat x$-coordinate of site $i$ and
\be
Q_Y =\sum_{i}^{y_1\geq 0} q_i.
\ee
\begin{definition}
\label{twAdef}
Let $H$ be any operator which can be written as $H=\sum_Z H_Z$ with the $H_Z$ supported on a set $Z$ of diameter less than
$L/2$.  Assume that all the sets $Z$ are squares.
Then, each such $H_Z$ intersects at most one of the lines $x=x_1$ or $x=x_2$ and at most one of the lines $y=y_1$ or $y=y_1$.
Then, define the {\bf twisted operator}
\be
H(\theta_1,\theta_2,\phi)
\ee
as follows.  
Let
\be
H(\theta_1,\theta_2,\phi)=\sum_Z H_Z(\theta_1,\theta_2,\phi),
\ee
where $H_Z(\theta_1,\theta_2,\phi)$ is defined as follows.
If the set $Z$ intersects
the vertical line $x=x_1$, then $H_Z(\theta_1,\theta_2,\phi)=\exp(i \theta_1 Q_X) H_Z(0,0,\phi) \exp(-i \theta_1 Q_X)$;
if the set $Z$ intersects
the vertical line $x=x_2$, then $H_Z(\theta_1,\theta_2,\phi_1)=\exp(-i \theta_2 Q_X) H_Z(0,0,\phi) \exp(i \theta_2 Q_X)$;
otherwise $H_Z(\theta_1,\theta_2,\phi_1,\phi_2)=H_Z(0,0,\phi)$.
Finally, define
$H_Z(0,0,\phi)=\exp(i \phi Q_Y) H_Z \exp(-i \phi Q_Y)$.
\end{definition}
Note that there is only one horizontal line for the annulus, while we had two for the torus.

\begin{figure}
\centering
\includegraphics[width=200px]{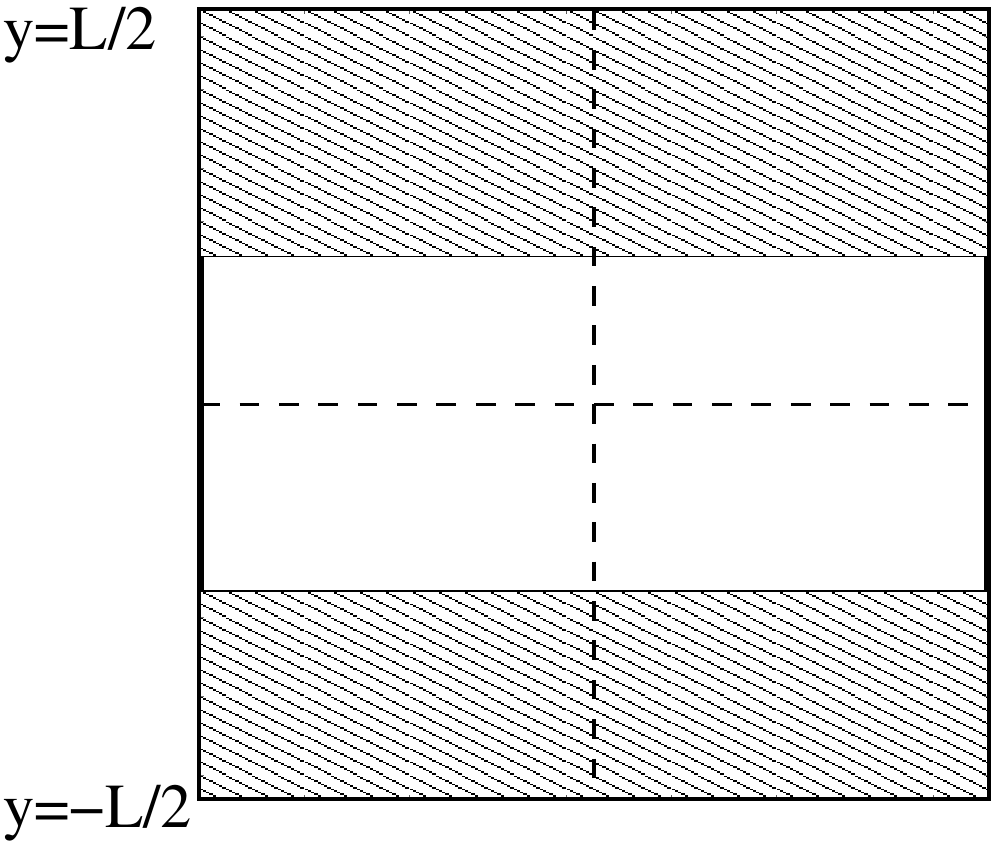}
\caption{{\small{Lines illustrating how the twists are defined on the annulus.  The twists $\theta_1,\theta_2$ affect interactions
close to the vertical solid and dashed lines, respectively, while the twist $\phi$ affects interactions
close to the horizontal dashed line.  Regions $\In$ and $\Out$ are the top and bottom rectangles filled with angled lines,
respectively.
}}}
\end{figure}

In this section, we prove the following:
\begin{theorem}
Consider a Hamiltonian with a bulk mobility gap, with $t_{max}$ superpolynomially large in $L$,
and $\lambda_{min}$ greater than or equal to $1/{\rm poly}(L)$, and
with
a $(c_0 L,\gub)$ unique bulk state, with $\gub$ greater than or equal to $1/{\rm poly}(L)$, for sufficiently small $c_0$.
Suppose we can find lines $x_1,x_2$ with $|x_1-x_2|=L/2$
such that the following holds. 
Let $A$ be the set of points within
distance $L/6$ of the intersection of the line $x=x_2$ with $y=0$.
Let $B$ be the union of the set of points within distance
$L/6$ of the other line and the
set of points within distance $L/6$ of either edge of the annulus.  Consider quasi-adiabatic evolution
under the quasi-adiabatic evolution operators defined below.
Suppose that $A$ and $B$ are separated under quasi-adiabatic evolution under all
such operators.
Then, the Hall conductance is quantized to $n(e^2/h)$,  for some integer $n$, up to an error which is superpolynomially small in $L$.
\end{theorem}

Note that we allow an operator on the line $x=x_1$ to evolve under quasi-adiabatic evolution to an operator which cannot be approximated by an operator
supported near $x=x_1$.  We allow the operator to spread out along the line,
and along the
boundary of the sample, having a large effect on sites along the boundary,
but we require that it not penetrate into the bulk near the line $x=x_2$.
This is needed for the second property below.

The proof essentially follows the above proof.
We again define a set of set of $N^2$ different closed paths in parameter space.  Each path keeps $\theta_2=\phi=0$ throughout.  Each
path is defined by a given pair of numbers $(\theta_x,\theta_y)$ with $\theta_x=m r$ and $\theta_y=n r$ for a pair of integers
$m,n$ in the range $0,...,N-1$. 
The paths start at $\theta_1=\phi=0$.  Then, we move to $\theta_1=0,\phi=\theta_y$.  Then we move to $\theta_1=\theta_x,\phi=\theta_y$.
Then we move around a small square loop of size $r$ at $\theta_x,\theta_y$ as follows: we move to
$\theta_1=\theta_x+r,\phi=\theta_y$, to
$\theta_1=\theta_x+r,\phi=\theta_y+r$, to
$\theta_1=\theta_x,\phi=\theta_y+r$, and back to
$\theta_1=\theta_x,\phi=\theta_y$.  Finally, we return to the origin by moving to
$\theta_1=0,\phi=\theta_y$ and then to $\theta_0=\phi=0$.

Thus, each path consists of $8$ different line segments.  We break it into three distinct parts.  First,
$V(m,n)$ moving from $0,0$ to 
to $\theta_1=0,\phi=\theta_y$.  Then we move to $\theta_1=\theta_x,\phi=\theta_y$.  Then,
$V_\circlearrowleft(m,n)$ moving around the small square loop.  Finally, $V^\dagger(m,n)$  returning to the origin.
We write the first part, $V(m,n)=U_2(m,n) U_1(m,n)$, with $U_{1}(m,n),U_2(m,n)$ corresponding to the two segments of
that part.  We also define unitary $\tilde U_1(m,n)$ and $\tilde U_2(m,n)$ corresponding to evolution along the
same path of  $\theta_1,\phi$ but with $\theta_2=-\theta_1$.

Given these paths, we have to pick a quasi-adiabatic evolution operator to generate the various unitaries $U$ and $\tilde U$.
We define ${\cal D}^{\theta}(\theta,\phi)$ to be the quasi-adiabatic
continuation operator which produces an infinitesimal
change in $\theta$ starting at $\theta_1=-\theta_2=\theta$ and the given $\phi$.  That is, for $H_s=H(\theta+s,-\theta-s,\phi)$, we
define ${\cal D}^{\theta}(\theta,\phi)={\cal D}_s$ at s=0.
We define ${\cal D}^{\phi}(\theta,\phi)$
to
produce an infinitesimal
change in $\phi$, again starting at $\theta_1=-\theta_2=\theta$ and the given $\phi$.  
We use these operators ${\cal D}^{\theta}(\theta,\phi),{\cal D}^{\phi}(\theta,\phi)$ to generate the unitaries
$\tilde U_1(m,n),\tilde U_2(m,n),\tilde U_{D,L,U,R}(m,n)$.
Let us consider the operator ${\cal D}^{\theta}(0,0)$.  For
$\theta_1=-\theta_2=s$ and $\phi=0$,
we have that $\partial_s H(s)=\sum_Z \partial_s H_Z(s)$, and
$\partial_s H_Z(s)$ is nonvanishing if $Z$ is within distance $R$ of the line at $x=x_1$ or if $Z$ is within
distance $R$ of the line at $x=x_2$.
Let $O^{(1)}(s)$ denote the sum of terms in $\partial_s H(s)$ near the line at $x=x_1$ and let
$O^{(2)}(s)$ denote the sum of terms in $\partial_s H(s)$ near the line at $x=x_2$, so that
${\cal D}_s={\cal D}(H_s,O^{(1)}(s))+
{\cal D}(H_s,O^{(2)}(s))$.
Under the assumption of a mobility gap, because of the superpolynomial localizability property,
up to superpolynomially small error. 
we can approximate the operators ${\cal D}(H_s,O^{(1)})$ and ${\cal D}(H_s,O^{(2)})$ by operators ${\cal D}_s^{(1)}$ and
${\cal D}_s^{(2)}$ supported within distance less than $L/8$ of the respective lines
$x=0$ and $x=L/2$ and which are a sum of squares of operators on squares of size at most $L/2$.
Then, we can apply the definition (\ref{twTdef}) to define 
\be
{\cal D}^{\theta,1}(\theta,\phi)={\cal D}_s^{(1)}(\theta,0,\phi),
\ee
and
\be
{\cal D}^{\theta,2}(\theta,\phi)={\cal D}_s^{(2)}(0,-\theta,\phi).
\ee
Note that
\be
\label{dcmptA}
{\cal D}^{\theta}(\theta,0)-
\Bigl( {\cal D}^{\theta,1}(\theta,0)
+{\cal D}^{\theta,2}(\theta,0) \Bigr)
\ee
is superpolynomially small.
Similarly, we define ${\cal D}^{\phi,1}$ as follows: we approximate ${\cal D}^{\phi}(0,0)$ by an operator ${cal D}_s$ which is
a sum of operators supported on squares of size at most $L/2$.  Then, we apply definition (\ref{twTdef}) to define
\be
{\cal D}^{\phi,1}(\theta,\phi)={\cal D}_s(\theta,0,\phi).
\ee
Note that ${\cal D}^{\phi}(\theta,\phi)$ is superpolynomially close to ${\cal D}_s(\theta,-\theta,\phi)$, so at $\theta\neq 0$,
the operators ${\cal D}^{\phi,1}(\theta,\phi)$ and ${\cal D}^{\phi}(\theta,\phi)$ are not necessarily superpolynomially close
to each other.
We will use these operators ${\cal D}^{\theta,1}(\theta,\phi)$ and ${\cal D}^{\phi,1}(\theta,\phi)$ to generate the unitaries
$U_1(m,n),U_2(m,n),U_{D,L,U,R}(m,n)$.

The proof rests on the following four facts.  We sketch each in turn.

First, for $r$ sufficiently small, the evolution around the path $U(0,0)$ returns the ground state to the ground state up to
superpolynomial small error and up to some phase which is
that is, up to superpolynomially
small error, equal to $r^2/2\pi$ times the Hall conductance in units of $e^2/h$, plus corrections of
order $r^3$.  
That is, similar to lemma (\ref{spcl}) above,
we claim that
\begin{lemma}
\label{spclA}
The quantity 
\be
\label{spcloA}
|\langle \Psi_0, U(0,0) \Psi_0 \rangle |
\ee
is superpolynomially close to unity for {\it any $r$}.  Also, the quantity
\be
\label{or3A}
|\langle \Psi_0, U(0,0) \Psi_0 \rangle - \exp(ir^2 \frac{\sigma_{xy}}{2\pi}\frac{h}{e^2})|
\ee
is bounded by terms of order ${\rm poly}(L) {\cal O}(r^3)$ plus terms
which are superpolynomially close to unity.
\end{lemma}
The proof of this is analogous to the torus case.  

The second property is that the evolution around all loops returns the ground state to the ground state and
produces the same phase up to superpolynomially small error.
To prove this, note that the desired expectation value is
$\langle \Psi_0 | V(m,n)^\dagger V_\circlearrowleft(m,n) V(m,n)| \Psi_0 \rangle$.  The operator
$V_{\circlearrowleft}$ can be approximated to superpolynomial accuracy by an
operator supported near $x=x_1,y=y_1$, using the Lieb-Robinson bound for quasi-adiabatic continuation.
So, 
$U_2(m,n) V_{\circlearrowleft}(m,n) U_2(m,n)$ is superpolynomially close to
$\tilde U_2(m,n)^\dagger V_{\circlearrowleft}(m,n) \tilde U_2(m,n)$.
However, $\tilde U_2(m,n)^\dagger  U_1(m,n) | \Psi_0\rangle=\exp(i Q_X \theta_1) \exp(i Q_Y\phi) | \Psi_0 \rangle$ for
$\theta_1=mr,\phi=nr$, and
$V_{\circlearrowleft}(m,n)$ is superpolynomially close to
$\exp(i Q_X \theta_1) \exp(i Q_Y\phi) 
V_{\circlearrowleft}(m,n)
\exp(-i Q_X \theta_1) \exp(-i Q_Y\phi)$.
So, 
$\langle \Psi_0 | U(m,n) | \Psi_0 \rangle $ is superpolynomially close to
$\langle \Psi_0 | U(0,0) | \Psi_0 \rangle $.
This argument is almost identical to the torus case, except different sets
were required to be
separated.

The third property is that the product $\Bigl(U(N-1,N-1) U(N-2,N-1) ... U(0,N-1)\Bigr) \Bigl( U(N-1,N-2) .... U(0,N-2) \Bigr)... \Bigl(U(N-1,0 ... U(0,0) \Bigr)$ is exactly equal to the unitary operator corresponding to quasi-adiabatic evolution around a big loop of size $2\pi$: moving from
$\theta_1=\phi_1=0$ to $\theta_1=2\pi,\phi_1=0$, to
$\theta_1=\phi_1=2\pi$ to $\theta_1=0,\phi_1=2\pi$, to $\theta_1=\phi_1=0$.  This is the evolution decomposition in \cite{hall}.

The fourth property is that the phase due to  evolution around the big loop is equal to unity up to superpolynomially small error.
To show this, note that evolution along the sides of the loop where $\phi$ increases from $0$ to $2\pi$, or decreases from $2\pi$ to zero,
is given by an operator that is, up to superpolynomially small error, supported in $\B$.  Define this operator
to be $U_{2\pi}^{\phi}$.  
Acting on the ground state, this operator $U^\phi_{2\pi}$
only produces a phase, since it approximates the exact evolution under a unitary transformation of the Hamiltonian.
This phase is non-trivial, being $z_\phi\equiv \exp(-i 2\pi\overline Q_Y)$, where $\overline Q_Y=\langle \Psi_0 | Q_Y| \Psi_0 \rangle$.

Let $U^\theta_{2\pi}$ define the evolution along the side of the big loop where $\theta$ increases from $0$ to $2\pi$.
We are interested in the action of $U^\phi_{2\pi}$ on the state $U^\theta_{2\pi}|\Psi_0\rangle$.  We claim that
$U^\phi_{2\pi} U^\theta_{2\pi}|\Psi_0\rangle \approx z_\phi U^{\theta}_{2\pi}|\Psi_0 \rangle$, and we will
show this in the next paragraph.  Given that this
is true, then the desired fourth property follows:
\be
\langle\Psi_0 |\Bigl( U^{\phi}_{2\pi} \Bigr)^\dagger \Bigl( U^\theta_{2\pi}\Bigr)^\dagger
U^\phi_{2\pi} U^\theta_{2\pi}|\Psi_0\rangle \approx 
\overline z_{\phi} z_\phi \langle \Psi_0 | 
\Bigl( U^\theta_{2\pi}\Bigr)^\dagger
U^\theta_{2\pi}|\Psi_0\rangle =1.
\ee

To show the claim, it suffices to show that the the reduced density matrix of
$U^\theta_{2\pi}|\Psi_0\rangle$ on $\B$ is close to the reduced density matrix of
$|\Psi_0\rangle$ on $\B$ since
$U^\theta_{2\pi}$ is approximated by an operator supported on $\B$.  To show this,  we use the assumption of a unique bulk state: if we can show that the expectation value in
the state
$U^\theta_{2\pi}|\Psi_0\rangle$
of all operators $O$ supported on
$\B$ supported on sets of diameter at most $c_0 L$ is close to that in the ground, then we are done.  If $O$ is distance
of order $L$ from
$x=x_1$, then this follows from the locality of the quasi-adiabatic continuation operator.  If $O$ is close to $x=x_1$,
then for sufficientlly small $c_0$, the operator $O$ is supported a distance of order $L$ from $x=x_2$ by assumption.
Define
\begin{eqnarray}
R_{2\pi}& \equiv &
{\cal S}' \exp\{i\int_0^{2\pi} {\rm d}s' {\cal D}_{s'}^{(1)}+{\cal D}_{s'}^{(2)} \} \\ \nonumber
& = &
U_{2\pi}^\theta
{\cal S}' \exp\Bigl\{i \int_0^{2\pi} {\rm d}s' (U_{s'}^\theta)^\dagger {\cal D}^{(2)} U_{s'}^{\theta}\Bigr\},
\end{eqnarray}
where $U_{s}^\theta=
{\cal S}' \exp\{i \int_0^{2\pi} {\rm d}s' {\cal D}_{s'}^{(1)}\}$.
Using the assumption of separation, the operator
${\cal S}' \exp\Bigl\{i \int_0^{2\pi} {\rm d}s' (U_{s'}^\theta)^\dagger {\cal D}^{(2)} U_{s'}^{\theta}\Bigr\}$ can be approximated to superpolynomial
accuracy by an operator supported away from the intersection of $\B$ with $x=x_1$.  So,
\be
\langle \Psi_0 |
\Bigl( U_{2\pi}^\theta \Bigr)^\dagger  O
U_{2\pi}^\theta  | \Psi_0 \rangle \approx 
\langle \Psi_0 |
R_{2\pi}^{\dagger}  O
R_{2\pi}  | \Psi_0 \rangle.
\ee
However, $R_{2\pi}$ describes quasi-adiabatic continuation along a unitarily equivalent path of Hamiltonians, and
$R_{2\pi}|\Psi_0\rangle=|\Psi_0\rangle$.  So, this completes the proof that the expectation value of $O$ is almost the same for
all operators $O$ of diameter at most $c_0 L$ supported in $\B$.

Given these four properties, we can complete in the same manner as in the torus case:
we approximate the phase of evolution around the large loop by the product of phases for evolution around each small loop.
The number of small loops scales as $1/r^2$, but
we can find an $r$ which is superpolynomially small but for which this leakage is negligible.
Hence, the product of the phases is superpolynomially close to unity.
Then, since the phase is almost  the same for every small loops, the phase around the loop with $m=n=0$ is close to
$2\pi n/N=2\pi(r/2\pi)^2$ for some integer $n$.  Hence the Hall conductance is close to $ne^2/h$.

\section{Discussion}
This paper mostly consisted of a series of definitions.  Once the correct definitions are found, the corrected quasi-adiabatic continuation
operator could be constructed straightforwardly.  This allowed us to carry over many of the previous results obtained with these operators in the case
of a spectral gap
in a straightforward way to the case of a mobility gap.  It is likely that there are other applications of these ideas.

In the appendix, we presented quasi-adiabatic continuation operators that have very rapid decay in space, simplifying and tightening
previous estimates.

{\bf Acknowledgments} I thank C. Nayak and T. J. Osborne for useful discussions on disordered systems.  I thank S. Bravyi and
S. Michalakis for useful discussions on filter functions.

\appendix
\section{Optimized Quasi-Adiabatic Continuation}
In this section, we construct an infinitely differentiable function $\tilde F(\omega)$ such that $\tilde F(\omega)=-1/\omega$ for
$|\omega|\geq 1$ and such that the Fourier transform $F(t)$ decays exponentially in a polynomial of $t$.  Our general strategy to construct
this function is to construct an infinitely differentiable function equal to $1$ for $|\omega|\geq 1$ and vanishing at $\omega=0$, and
then multiply this function by $1/\omega$.

Then, in the next section we show the application of this function to construct an exact (in that it can be used
to exactly describe evolution of the ground state of a gapped Hamiltonian under a parameter change) quasi-adiabatic operator with good decay
properties in space (we present this in detail only for a system with a spectral gap), and show Lieb-Robinson bounds for this operator and
discuss the best possible bounds.
Then, we present some discussion of the problem of optimizing over functions.
Finally, we present applications in the last section.

Let $\tilde f(\omega)$ be even and have the property that
$\tilde f(\omega)=0$ for $\omega=0$ and $\tilde f(\omega)=1$ for $|\omega|\geq 1$.
Let $f(t)$ be the Fourier transform of $\tilde f(\omega)$.
Then, define
$F(t)$ by
\be
\label{da}
F(t)=\frac{i}{2}\int {\rm d}u f(u) \sgn(t-u),
\ee
where $\sgn(t-u)$ is the sign function: $\sgn(t-u)=1$ for $t>u$, $\sgn(t-u)=-1$ for $t<u$, and $\sgn(0)=0$.
We now show the time decay of $F(t)$ and we show that
the Fourier transform $\tilde F(\omega)$ is equal to $-1/\omega$ for $|\omega|\geq 1$, as desired. 
\begin{lemma}
Let $F(t)$ be as defined in \ref{da}.  Let $\tilde F(\omega)$ be the Fourier transform of $F(t)$.  Then,
\be
|F(t)|\leq |\int_{|t|}^{\infty} f(u)  {\rm d}u|,
\ee
and
\be
\tilde F(\omega)=
\frac{-1}{\omega} \tilde f(\omega).
\ee
\begin{proof}
Assume, without loss of generality, that $t\geq 0$.  Then, we have
$|F(t)|\leq |\int_{t}^{\infty} f(u)  {\rm d}u|/2+
|\int_{-\infty}^{t} f(u) {\rm d}u|/2$.  Since $\tilde f(0)=0$, we have
$|\int_{-\infty}^{t} f(u) {\rm d}u|=
|\int_{t}^{\infty} f(u)  {\rm d}u|$.  Thus,
$|F(t)|\leq |\int_{t}^{\infty} f(u)  {\rm d}u|$.

We have
\be
\tilde F(\omega)=
\frac{i}{2}
\int {\rm d}t \exp(i \omega t) 
\int {\rm d}u f(u) \sgn(t-u).
\ee
Integrating by parts in $t$, we have
\begin{eqnarray}
\tilde F(\omega)&= &
\frac{-1}{\omega} \int {\rm d}t \exp(i \omega t) 
\int {\rm d}u f(u) \delta(t-u) \\ \nonumber
&=&
\frac{-1}{\omega} \tilde f(\omega).
\end{eqnarray}
Note that $\lim_{t\rightarrow \pm \infty} 
\Bigl( \int {\rm d}u f(u) \sgn(t-u)\Bigr)=0$, so the contributions to the integration by parts from the upper and lower limits
of integration vanish.
\end{proof}
\end{lemma}

Thus, given any function $f(t)$ which decays as $|f(t)| \leq C \exp(-t^\alpha)$, for some $C,\alpha>0$, we can
find a filter function $F(t)$ which decays as $C \exp(-t^\alpha)$, for some other $C$.
Note that given any even function
$g(t)$ which has the Fourier transform with the property that
$\tilde g(\omega)=1$ for $\omega=0$, $\tilde f(\omega)=0$ for $|\omega|\geq 1$, we can
define
\be
f(t)=\delta(t)-g(t),
\ee
where $\delta(t)$ is the Dirac $\delta$-function (since we convolve $f(u)$ against $\sgn(t-u)$, the resulting $F(t)$ is
a function, rather than a distribution).

In the classic paper\cite{note}, it is shown how to construct such functions $g(t)$ such
that
\be
|g(t)|\leq {\cal O}(\exp(-|t| \epsilon(|t|))),
\ee
for {\it any} monotonically decreasing positive function $\epsilon(y)$ such that
\be
\int_1^{\infty} \frac{\epsilon(y)}{y} {\rm d}y 
\ee
is convergent.
For example, such a function $\epsilon(y)$ may be chosen to be
\be
\epsilon(y)=1/\log(2+y)^2.
\ee
Thus, our function $g(t)$ has
so-called ``subexponential decay"\cite{subexp}.  A function $f(t)$
is defined to have subexponential decay if,
for any $\alpha<1$,
$|f(t)| \leq C_{\alpha} \exp(-t^\alpha)$, for some $C_{\alpha}$ which depends on $\alpha$.
Thus,
\begin{corollary}
There exist odd functions $F(t)$ with subexponential decay and with $\tilde F(\omega)=-1/\omega$ for $|\omega|\geq 1$.
In fact, the resulting function $F(t)$ from the given $\epsilon(y)$ obeys
\be
\label{infront}
|F(t)| \leq {\cal O}(\log(2+|t|)^2 \exp(-|t|/\log(2+|t|)^2)).
\ee
\end{corollary}
Below Eq.~(\ref{gfnlist}), we list even faster possible asymptotic decay.

\section{Decay of Quasi-Adiabatic Continuation Operator}
Given the rapid decay of $F(t)$ with $t$, we have a Lieb-Robinson bound for quasi-adiabatic continuation using this operator, as we now
show.
Assume a system has a spectral gap $\gamma$, and that $H_s=\sum_Z H_s(Z)$, with $H_s(Z)$ supported on sets $Z$ of diameter at most
$R$.
Define a quasi-adiabatic evolution operator
by
\be
{\cal D}_s=\sum_Z {\cal D}^Z_s,
\ee
where 
\be
i{\cal D}^Z_s=\int {\rm d}t F(\gamma t) \exp(i H_s t) \Bigl(\partial_s H_s(Z) \Bigr) \exp(-iH_st).
\ee
If a Hamiltonian $H_s$ has a spectral gap $\gamma$ and a ground state $\Psi_0(s)$, then by construction
\be
\partial_s \Psi_0(s)=i{\cal D}_s \Psi_0(s).
\ee

We now consider the decay properties of $i{\cal D}^Z_s$.  Let $O$ be an operator supported on a set which is distance
at least $l$ from $Z$.  Then, using  the Lieb-Robinson bound and a triangle inequality
\begin{eqnarray}
\Vert[O,{\cal D}^Z_s]\Vert &\leq &\int {\rm d}t |F(\gamma t)| \Vert [O,\exp(i H_s t) \Bigl(\partial_s H_s(Z) \Bigr) \exp(-iH_st)] \Vert
\\ \nonumber
&\leq & 
\Bigl( \int_{|t|\leq l/\vlr} {\rm d}t |F(\gamma t)| g(l) +
2\int_{|t|\geq l/\vlr} {\rm d}t |F(\gamma t)| \Bigr) \times
\Vert \partial_s H_s(Z) \Vert \Vert O \Vert
\\ \nonumber
&\leq &
\EE(l) \Vert \partial_s H_s(Z) \Vert \Vert O \Vert,
\end{eqnarray}
where we define
\be
\EE(l)\equiv
\frac{1}{\gamma} \Bigl( \int {\rm d}u |F(t)| g(l)+2\int_{u\geq l\gamma/\vlr} |F(u)| \Bigr),
\ee
where we used the change of variables, $u=\gamma t$.  We define, for use later, $\EE(0)=1$.

Define $b_l(Z)$ to be the set of sites within distance $l$ of $Z$.
Define an approximation to ${\cal D}^Z_s$ which is supported on $b_l(Z)$ by
\be
{\cal D}^{Z,l}_s=\int {\rm d}U U^{\dagger} {\cal D}^Z_s U,
\ee
where the integral ranges over unitaries $U$ which are suspported on the complement of $b_l(Z)$ with the 
Haar measure.  Using  the commutator estimate
above, we can estimate the difference $\Vert U^\dagger {\cal D}^Z_s U-{\cal D}^Z_s \Vert \leq \Vert [U,{\cal D}^Z_s] \Vert$.
So
\be
\Vert {\cal D}^{Z,l}_s-{\cal D}^Z_s\Vert \leq  \EE(l).
\ee

Define a sequence of operators $D_s(Z,j)$ by
\be
D_s(Z,0)={\cal D}^{Z,0}_s,
\ee
for $j=0$ and by
\be
D_s(Z,j)={\cal D}^{Z,j}_s-{\cal D}^{Z,j-1}_s,
\ee
for $j>0$.
We have the bound for $j>0$, that $\Vert D_s(Z,j) \Vert \leq 2 \EE(l)$.
Thus, since
\be
{\cal D}_s=\sum_{Z,j} D_s(Z,j),
\ee
we have decomposed ${\cal D}_s$ as a sum of operators on sets of increasing size and decreasing norm as follows.

For definiteness, let us assume that the initial Hamiltonian $H$ has the property that
the sets $Z$ are all given by balls of radius $R/2$ about sets $i$ (let $R$ be even for simplicity).  So, we write $H_Z=H_i$, where $Z=b_{r/2}(\{i\})$, and
$\{i\}$ is the set containing just site $i$.  Then,
\begin{lemma}
For the given function $F(t)$,
\be
{\cal D}_s=\sum_{i\in \Lambda} \sum_{j\geq R/2} D_s(i,j),
\ee
where the sum is over sites $i$ in the lattice $\Lambda$, and
where $D_s(i,j)$ is an operator supported on $b_j(\{i\})$ and
\be
\Vert D_s(i,j) \Vert \leq 2\EE(l-R/2) \Vert \partial_s H_Z(s) \Vert.
\ee
\end{lemma}

If a set $Y=b_j(\{i\})$, then we define
$D_s(Y)=D_s(i,j)$.
It is also worth bounding this as:
\begin{lemma}
\label{Know}
For any pair of sites $i,j$,
\be
\sum_{Y\ni i,j} \Vert D_s(Y) \Vert \leq K({\rm dist}(i,j)),
\ee
where
\be
K(l)\leq 
J'
\sum_{m\in \Lambda}\; \sum_{k\geq {\rm max}({\rm dist}(i,m),{\rm dist}(j,m))} \;  \EE(k).
\ee
where we define
\be
J'={\rm max}_Z (\Vert \partial_s H_Z(s) \Vert).
\ee
\end{lemma}

We are now ready to derive the Lieb-Robinson bounds for evolution under
${\cal D}_s$.
Define a unitary $U_s$ corresponding to evolution with ${\cal D}_s$ by
\be
U_s={\cal S}' \exp\{i\int_0^s {\rm d}s' {\cal D}_{s'}\}.
\ee
Let $O_A$ be any operator supported on set $A$, and let $O_B$ be an operator supported on set $B$, where $B$ is the set of
sites which are distance at least $l$ from set $A$.
We wish to bound
\be
\Vert [ U_s O_A U_s^\dagger,O_B ] \Vert.
\ee

If a set $Y=b_{j}(\{i\})$, we will write $D_s(Y)=D_s(i,j)$ to save notation in what follows.
We use the series expansion first derived in \cite{lr2}:
\begin{eqnarray}
\Vert [ U_s O_A U_s^\dagger,O_B ] \Vert & \leq  &
2 (2|s|) \sum_{Y:Y\cap A\neq \emptyset,Y\cap B \neq \emptyset} \Vert D_s(Y) \Vert \\ \nonumber
&& +
2 \frac{(2|s|)^2}{2!} \sum_{Y_1,Y_2:Y\cap A\neq \emptyset,Y_1\cap Y_2 \neq \emptyset, Y_2\cap B \neq \emptyset} 
\Vert D_s(Y_1) \Vert 
\Vert D_s(Y_2) \Vert 
\\ \nonumber
&&+
2 \frac{(2|s|)^3}{3!} \sum_{Y_1,Y_2,Y_3:Y_1\cap A\neq \emptyset,Y_1\cap Y_2\neq \emptyset,Y_2\cap Y_3\neq\emptyset, Y_3\cap B \neq \emptyset} 
\Vert D_s(Y_1) \Vert
\Vert D_s(Y_2) \Vert
\Vert D_s(Y_3) \Vert
 \\ \nonumber
&&+...
\end{eqnarray}

We may bound this as follows:
\be
\Vert [ U_s O_A U_s^\dagger,O_B ] \Vert \leq \sum_{i\in A} \sum{j\in B} G(i,j),
\ee
where
$G(i,j)$ is defined by
\begin{eqnarray}
\label{Gijdef}
&& 2 (2|s|) \sum_{Y:i \in Y,j\in Y} \Vert D_s(Y) \Vert \\ \nonumber
&& +
2 \frac{(2|s|)^2}{2!} \sum_{Y_1,Y_2:i \in Y,Y_1\cap Y_2 \neq \emptyset, j\in Y_2} 
\Vert D_s(Y_1) \Vert 
\Vert D_s(Y_2) \Vert 
\\ \nonumber
&&+
2 \frac{(2|s|)^3}{3!} \sum_{Y_1,Y_2,Y_3:i\in Y_1,Y_1\cap Y_2\neq \emptyset,Y_2\cap Y_3\neq\emptyset, j\in Y_3}
 \Vert D_s(Y_1) \Vert
 \Vert D_s(Y_2) \Vert
 \Vert D_s(Y_3) \Vert
 \\ \nonumber
&&+...
\end{eqnarray}

We define 
\begin{definition}
A function $K(l)$ is {\bf reproducing} for a given lattice $\Lambda$ if, for any pair of sites $i,j$ we have
\be
\sum_m K({\rm dist}(i,m)) K({\rm dist}(m,j) \leq \lambda K({\rm dist}(i,j)),
\ee
for some constant $\lambda$.
\end{definition}
For a square lattice in $D$ dimensions and a shortest-path metric, a powerlaw $K(l)\sim l^{-\alpha}$ is reproducing for sufficiently
large $\alpha$.  An exponential decay is {\it not} reproducing.  However an exponential multiplying a sufficiently fast decaying power
is.  Using this definition and lemma (\ref{Know}) and Eq.~(\ref{Gijdef}), we arrive at the bound for a reproducing $K(l)$ that
\begin{eqnarray}
G(i,j) & \leq & 2 K({\rm dist}(i,j)) (2|s|+\frac{(2|s|)^2}{2!} \lambda + 
\frac{(2|s|)^3}{3!} \lambda^2 + ... \\ \nonumber
&\leq & 2 
K({\rm dist}(i,j)) \frac{\exp(2\lambda |s|)-1}{\lambda}.
\end{eqnarray}

Now, we consider a specific case.
In order to maintain generality in our choice of $F(t)$, we want one more definition.
\begin{definition}
A function $f(x)$ is of {\bf decay class} $g(x)$ if
\be
f(x) \leq c_1 g(c_2 x)
\ee
for all $x\geq 0$, for some constants $c_1,c_2$.  
\end{definition}
In some cases we will specify the constants.

Consider a function $g(t)$ defined to have 
$|g(t)|\leq {\cal O}(\exp(-|t| \epsilon(|t|)))$, as constructed in \cite{note}.  Suppose
$\epsilon(t)$ is any one of the following functions:
\be
\label{gfnlist}
\frac{C}{\log(t)^2}\; ,\; \frac{C}{\log(t)\log(\log(t))^2} \;, \;
\frac{C}{\log(t)\log(\log(t))\log(\log(\log(t)))^2},...
\ee
From any such function, in the first section of the appendix we succeed in constructing a function $F(t)$ which is
of decay class $g$.  One may note that Eq.~(\ref{infront})
has the extra factor $\log(2+|t|)^2$ multiplying the decay of $F(t)$, but this is still asymptotically bounded by
${\cal O}(\exp(-c_2 |t|/\log(2+|t|)^2))$, for any $c_2<1$, so indeed $F(t)$ is of decay class $g(t)$.

We now restrict to finite dimensional lattices, so that the number of sites within distance $l$ of any site grows
at most polynomially with $l$.  Then, we find that the function $\EE$ is also of decay class $g$.  However, here
it is important to specify the constants: the constant $c_1$ is proportional to $1/\gamma$, while the constant
$c_2$ is proportional to $\gamma/\vlr$.
The function $K$ is of decay class $\EE$ with constants $c_1,c_2$ that depend on the dimension: the number of sites $m$ in the
sum in Eq.~(\ref{Know}) will be dimension-dependent.  We are not interested in these dimension dependent constants, but only
in the constants that depend on $\gamma$.  Note that $K$ is of decay class $g$ with
$c_1$ proportional to $J'/\gamma$ and
$c_2$ proportional to $\gamma/\vlr$.

The functions $K$ which arise from these functions $g$ decay faster than $1/x^\alpha$ for any power $\alpha$.
So, for any $\alpha>0$, for any $c_2<1$, $K(x)$ can be upper bounded by 
\be
K(x) \leq K'(x)/x^\alpha,
\ee
where $K'(x)=c_1 K(c_2 x)$ for some $c_1$.  Note that for sufficiently large $\alpha$ such functions $K'$ are reproducing, and indeed one can pick $\lambda \leq
C (J'/\gamma)$, for a dimension-dependent constant $C$ of order unity.
Also,  $K'$ is of decay class $K$.
Define
\be
G(l)={\rm max}_{i,j}^{{\rm dist}(i,j)=l} G(i,j),
\ee
so that $G(l)$ is an upper bound on $G(i,j)$ if ${\rm dist}(i,j)=l$.  Then,
\begin{lemma}
For
any of the functions $g(t)$ arising from $\epsilon$ in Eq.~(\ref{gfnlist}), the corresponding
$G(l)$ is bounded by  $\exp(C|s| J'/\gamma)-1$ times $c_1 g(c_2 l \gamma/\vlr)$, for some constant $c_1,c_2$ which
are dimension dependent, but do not depend on $\gamma$ or $\vlr$, and
\end{lemma}

Similarly, in any finite dimensional lattice, after we sum over sites $i,j$ we find that
\begin{lemma}
\label{leaklemma}
For
any of the functions $g(t)$ arising from $\epsilon$ in Eq.~(\ref{gfnlist}), 
the commutator
$\Vert [ U_s O_A U_s^\dagger,O_B ] \Vert$ is bounded by $|A|$ times
by  $\exp(C|s| J'/\gamma)-1$ times $c'_1 g(c'_2 l \gamma/\vlr)$, for some constants $c'_1,c'_2$ which
are dimension dependent, but do not depend on $\gamma$ or $\vlr$.
\end{lemma}

\section{Optimization}
In the next section, we will apply this lemma (\ref{leaklemma}) to specific problems, where the error estimates depend upon
the error in lemma (\ref{leaklemma}) for $l$ of order the system size $L$.
Hence, we will find errors that
decay with system size $L$ similarly to the decay of $g(t)$ with $t$.
In this section, we discuss optimizing these error estimates in a little more detail; the reader may prefer to skip this until later.
The reader will perhaps realize that instead of choosing a function, such a $\exp(-|t|/\log^2(t))$, we could easily have chosen the function
$\exp(-2|t|/\log^2(t))$ or indeed
$\exp(-C|t|/\log^2(t))$ for any constant $C$.  One may then start to get optimistic: if, for example, we prove that a certain
error decays as $\exp(-1000 L/\log^2(L))$, surely this is much better than proving that it decays only as
$\exp(-L/\log^2(L))$.  Of course, the trouble is the constant in front that we have not written: all the of the bounds on the decay of $g(t)$ are
up to a constant in front, and by choosing a $g(t)$ which decays asymptotically
as ${\cal O}(\exp(-1000 |t|/\log^2(t)))$, we will find a much worse constant than if we had chosen $g(t)$ to decay only as
as ${\cal O}(\exp(-|t|/\log^2(t)))$.  In fact, one will find that one gains only for very large $t$, and hence, for very large $L$.

The next possibility one may consider is:
suppose, for definiteness, we choose $g(t)$ to decay as $\exp(-C|t|/\log^2(t))$ for a constant $C$.  How can we optimize the constant $C$ to obtain
the best decay for a given system size $L$?  After all, perhaps if $L$ is large one should choose a large constant $C$ in defining
$g(t)$, while for smaller $L$ perhaps a smaller constant is better.  However,
as soon as one starts trying to optimize over constants, one should also try to optimize over functions.
In order to obtain the best possible bounds, one would like to be able to consider all possible $g(t)$ for a given $L$.
Thus, the problem which one would really like to solve is:
for a given distance $L$, and a given dimension $D$, what function $g(t)$ will minimize 
the quantity $\exp(C|s| J'/\gamma)-1$ times $c'_1 g(c'_2 L \gamma/\vlr)$ in lemma (\ref{leaklemma}) for $|s|$ of order
unity?
In particular, how does the minimum over $g(t)$ of this quantity behave for large $L$?
It should be clear that the term
$\exp(C|s| J'/\gamma)-1$ is important.  While $|s|$ is of order unity, if $C$ becomes large (for example, $L$-dependent) then this
term will become large.

\section{Applications}
These operators immediately simplify and tighten several results.  For example, 
\cite{short,long} rely on quasi-adiabatic continuation operators; using the operator here, the superpolynomial decay of the
ground state splitting in those
papers can be tightened to subexponential.

Further, these operators can be directly inserted into the proof  in \cite{hall}.  This greatly simplifies the construction
of the quasi-adiabatic evolution operator.  Further, one immediately finds that 
\begin{theorem}
Let $g(t)=\exp(-t\epsilon(t))$, where
$\epsilon(t)$ is any function in Eq.~(\ref{gfnlist}).
For Hamiltonians on a torus, as considered in \cite{hall}, for fixed $R,J/\gamma$, the Hall conductance is quantized to an integer up
to an error
which is bounded by a function $e(L)$, where $e$ is of decay class $g$. 
\end{theorem}


\begin{thebibliography}{99}
\bibitem{ak}  While this work was in progress, a similar definition was mentioned to me by A. Kitaev, under the name ``locally definable".
\end{thebibliography}
\end{document}